\documentclass[12pt ]{article}
\usepackage{amsmath}
\usepackage{graphicx}
\usepackage{hyperref}
\usepackage{lmodern}
\usepackage{amssymb}
\usepackage{booktabs}
\usepackage{tikz}
\usetikzlibrary{patterns}

\renewcommand{\theequation}
{\arabic{section}.\arabic{equation}}

\makeatletter
\def\eqnarray{ \stepcounter{equation} \let\@currentlabel=\theequation
 \global\@eqnswtrue
 \global\@eqcnt\z@
 \tabskip\@centering
 \let\\=\@eqncr
 $$\halign to \displaywidth\bgroup\@eqnsel\hskip\@centering
 $\displaystyle\tabskip\z@{##}$&\global\@eqcnt\@ne
 \hfil$\displaystyle{{}##{}}$\hfil
 &\global\@eqcnt\tw@$\displaystyle\tabskip\z@{##}$\hfil
 \tabskip\@centering&\llap{##}\tabskip\z@\cr}
\makeatother

\newcommand{\be}{\begin{equation}}
\newcommand{\ee}{\end{equation}}

\newcommand{\beqa}{\begin{eqnarray}}
\newcommand{\eeqa}{\end{eqnarray}}
\newcommand{\nn}{\nonumber}

\def\CB {{\cal B}}

\def\CE {{\cal E}}
\def\CF {{\cal F}}
\def\CG {{\cal G}}
\def\CH {{\cal H}}
\def\CI {{\cal I}}

\def\CL {{\cal L}}

\def\CR {{\cal R}}
\def\CS {{\cal S}}

\begin{document}

\setlength{\baselineskip}{7mm}
\begin{titlepage}
\begin{flushright}

{\tt NRCPS-HE-01-2024} 
\end{flushright}

\begin{center}
{\Large ~\\{\it   How Large is the Space of Covariantly  Constant Gauge Fields
}

}

\vspace{2cm}

{\sl George Savvidy

\centerline{${}$ \sl Institute of Nuclear and Particle Physics}
\centerline{${}$ \sl Demokritos National Research Center, Ag. Paraskevi,  Athens, Greece}

}
 
\end{center}
\vspace{2cm}

\centerline{{\bf Abstract}}

The covariantly constant gauge fields are solutions of the sourceless Yang-Mills equation and represent the classical vacuum fields. We found that the moduli space of covariantly constant gauge fields is much larger than the space of constant chromomagnetic fields.  A wider class of  covariantly constant gauge field solutions representing non-perturbative magnetic flux sheets of finite thickness is obtained through the nontrivial space-time dependence of a unit colour vector. In some sense these solutions are similar to the Nielsen-Olesen magnetic flux tubes, but instead they have geometry of magnetic flux sheets and are supported without presence of any Higgs field.  The infinitesimally thin sheet solutions can be associated  with the  singular surfaces considered by 't Hooft.  The nonlocal operators that are supported on a two-dimensional surface rather than a one-dimensional curve where considered in literature. These surface operators are analogous to Wilson $W(C)$ and ’t Hooft line operators $M(C)$ except that they are supported on a two-dimensional surface rather than a one-dimensional curve. The class of non-perturbative solutions representing a nonvanishing chromomagnetic field that fills out the whole 3D-space is obtained as well. This  new class  of covariantly constant gauge field solutions is constructed by using the general properties of the Cho Ansatz.  We define the topological currents and the corresponding  charges and demonstrate that the new solutions have a zero monopole charge density. Instead, the solutions have a nonzero Hopf invariant, which is the magnetic helicity of the Faraday force lines.

  \vspace{12pt}

\noindent

\end{titlepage}

\pagestyle{plain}

\section{\it Introduction}

The investigation of the Yang-Mills vacuum polarisation    \cite{Batalin:1976uv, Savvidy:1977as, Matinyan:1976mp} revealed that the effective Lagrangian of the SU(N) Yang-Mills  theory  has the following form:
\beqa\label{YMeffective}
\CL  =  
-\CF - {11  N \over 96 \pi^2}  g^2 \CF \Big( \ln {2 g^2 \CF \over \mu^4}- 1\Big),
\eeqa
where  $\CF= {1\over 4} G^a_{\mu\nu}G^a_{\mu\nu} ={\vec{\CH}^2_a -\vec{\CE}^2_a\over 2} >0$ and   $\CG = {1\over 4} G^a_{\mu\nu} \tilde{G}^a_{\mu\nu} = \vec{\CH}_a \vec{\CE}_a =0$, and that the vacuum energy density has its new minimum  at a nonzero value of the field strength:
\be\label{chomomagneticcondensate}
 \langle 2  g^2 \CF \rangle_{vac}=    \mu^4  \exp{(-{32 \pi^2 \over 11 g^2(\mu) })} = \Lambda^4_{S}.
\ee
The chromomagnetic magnetic induction $\vec{\CB}_a$  of the YM vacuum  is given by the formula  \cite{PhDTheses}:
\be\label{palarYM}
\vec{\CB}_a = - {\partial \CL \over \partial \vec{\CH}_a} 
= \vec{\CH}_a ~ {11 g^2  N \over 96 \pi^2}\log{g^2 \vec{\CH}^2_a \over \Lambda^4_{S}} ~= \mu_{vac}~   \vec{\CH}_a.
\ee
 The paramagnetism of the YM vacuum at  $g^2 \vec{\CH}^2_a \geq \Lambda^4_{S}$ means that there is an amplification of the chromomagnetic vacuum fields similar to the Pauli paramagnetism, an effect associated with the polarisation of the virtual vector boson spins.     

In the earlier investigations of the vacuum polarisation by covariantly constant gauge fields it was realised  that the consideration of the vacuum polarisation in the quadratic  approximation \cite{Batalin:1976uv, tHooft:1976snw} displays an apparent instability of the chromomagnetyic field  due to the negative/unstable mode  and the appearance of an imaginary term in the effective Lagrangian \cite{Nielsen:1978rm, Skalozub:1978fy, Ambjorn:1978ff, Nielsen:1978zg,  Ambjorn:1980ms,   Leutwyler:1980ev, Leutwyler:1980ma, Minkowski:1981ma, Flory:1983td, Faddeev:2001dda, Savvidy:2019grj}.   

Generalising  the calculation advocated earlier  by Ambjorn, Nielsen, Olesen \cite{ Ambjorn:1978ff, Nielsen:1978zg, Ambjorn:1980ms},  Flory \cite{Flory:1983td},  and other authors \cite{ Leutwyler:1980ev, Leutwyler:1980ma,  Minkowski:1981ma, Faddeev:2001dda,  Pimentel:2018nkl,  Savvidy:2019grj, Parthasarathy:1983ck, Kay:1983an, Kay1983, Dittrich:1983ej, Zwanziger:1982na, Kay:2005wm, Kondo:2013aga, Cho:2004qf, Pak:2020izt, Pak:2020obo, Pak:2017skw, Pak:2020fkt, Baseian:1979zx} one can perform  an exact  integration over {\it nonlinearly  interacting negative/unstable modes}  and  obtain the effective Lagrangian, which does not contain imaginary terms  and is identical to the expression (\ref{YMeffective}) \cite{Savvidy:2022jcr, Savvidy:2023kmx, Savvidy:2023kft}. This consideration reflects the fact that a magnetic field does not produce work  and cannot create particle pairs from the vacuum \cite{Savvidy:1977as},  opposite to what takes place in the case of chromoelectric fields.  The  covariantly constant  chromomagnetic vacuum fields   proved to be  stable and indicated that {\it the Yang-Mills vacuum is a highly degenerate  quantum state} \cite{Savvidy:2022jcr, Savvidy:2023kmx, Savvidy:2023kft}.

In this respect it seems important to investigate a {\it moduli space of vacuum field configurations} and describe precisely the degeneracy of the classical vacuum fields.  An early attempt to find a larger class of  space-homogeneous vacuum  Yang-Mills fields was made in \cite{Baseian:1979zx, Savvidy:1982wx, Savvidy:1982jk, Matinyan:1981ys, Matinyan:1981dj,Banks:1996vh}.  It was shown that space-homogeneous vacuum fields exhibit a deterministic  chaos \cite{Savvidy:2020mco}. The vacuum fields were also considered in \cite{ Cho:2004qf, Pak:2020izt, Pak:2020obo, Pak:2017skw, Pak:2020fkt, Nielsen:1979vb,  Anous:2017mwr, Cho:1979nv, Kim:2016xdn, Milshtein:1983th, Olesen:1981zp,  Apenko:1982tj, Reuter:1994yq, Reuter:1997gx, Reuter:1994zn, Savvidy:2020mco, Wu:1975vq, Wu:1967vp}. The class of solutions  in the background field  was investigated in a series of articles  and has a "spaghetti" type structure of magnetic tubes \cite{Nielsen:1978tr,Nielsen:1979xu, Ambjorn:1979xi,Ambjorn:1980ms}.

The problem that remained unsolved was to identify {\it how large is the class of  covariantly constant gauge fields} defined by the equation   \cite{Batalin:1976uv, Savvidy:1977as, Matinyan:1976mp, Brown:1975bc, Duff:1975ue} 
\be
\nabla^{ab}_{\rho} G^{b}_{\mu\nu} =0.
\ee
The covariantly constant gauge fields represent a subclass of solutions of the sourceless Yang-Mills equation  
\be\label{YMeq} 
\nabla^{ab}_{\mu} G^{b}_{\mu\nu} =0.
\ee 
The main property of covariantly constant gauge fields is that the Lorentz and colour structures factorise  \cite{Batalin:1976uv, Savvidy:1977as, Matinyan:1976mp, Brown:1975bc, Duff:1975ue}:
\be\label{factor} 
G^{a}_{\mu\nu}(x)= G_{\mu\nu}(x) n^a(x) .
\ee
The  importance of the covariantly constant gauge fields consists in the fact that they are the solutions of the sourceless vacuum Yang-Mills equation and that {\it the effective Lagrangian is gauge invariant only on sourceless vacuum fields} \cite{Batalin:1976uv,Batalin:1979jh}. We will show that the class of covariantly constant gauge fields is much larger than the fields defined by the equation  
$$
A^{a}_{\mu} = - {1\over 2} F_{\mu\nu} x_{\nu}   n^a,
$$
where $n^a$ is a constant colour vector.  We will demonstrate that a  wider class of covariantly-constant gauge fields can be obtained through the nontrivial space-time dependence of the unit vector $n^a(x)$ and that it generates an additional contribution $S_{\mu\nu}$ to the field strength tensor. 
This  new class  of covariantly constant gauge fields can be constructed by using the general properties of the Cho Ansatz  \cite{Cho:1979nv, Cho:1980nx}:
$$
A^{a}_{\mu} =  B_{\mu} n^{a}  +
{1\over g} \varepsilon^{abc} n^{b} \partial_{\mu}n^{c}.
$$
One  of the important properties of the Cho Ansatz is that the corresponding field strength tensor factorises into the Lorentz and colour  structures:
$$
G^{a}_{\mu\nu} =  ( F_{\mu\nu} + {1\over g} S_{\mu\nu})~ n^{a}(x),
$$ 
very similarly to the factorisation  (\ref{factor})  in the case of covariantly constant gauge fields. It is therefore natural to search for new solutions  in the form of  Cho Ansatz.  The new class of  trigonometric solutions is defined in the whole space-time and polynomial solutions  representing non-perturbative magnetic sheets of finite thickness will be derived in the forthcoming  sections.   The new solutions have zero monopole charge. Instead of that the solutions have a nonzero Hopf-Chern-Simon topological density distributed over the whole space.

\section{\it Covariantly Constant Gauge Fields }

The covariantly-constant gauge fields are defined by the equation \cite{Batalin:1976uv, Savvidy:1977as, Matinyan:1976mp, Brown:1975bc, Duff:1975ue}
\be\label{YMeqcov}
\nabla^{ab}_{\rho} G^{b}_{\mu\nu} =0,
\ee
where 
\be\label{fieldstregth}
G^{a}_{\mu\nu} =  \partial_{\mu} A^{a}_{\nu} - \partial_{\nu} A^{a}_{\mu}
 - g \varepsilon^{abc} A^{b}_{\mu} A^{c}_{\nu}
\ee
and 
\be
\nabla^{ab}_{\mu}(A)=
  \delta^{ab} \partial_{\mu}   - g \varepsilon^{acb} A^{c}_{\mu} .
\ee
It  follows that the covariantly constant gauge fields are solutions of the vacuum Yang-Mills equation (\ref{YMeq}). By taking covariant derivative $\nabla^{ca}_{\lambda}$ of the l.h.s and interchanging the derivatives one can get 
\be\label{comzero}
[\nabla_{\lambda} , \nabla_{\rho}]^{ab} G^{b}_{\mu\nu} =0.
\ee
The commutator of covariant derivatives is equal to the field strength  tensor:
\be
[\nabla_{\lambda} , \nabla_{\rho}]^{ab} = -g  \varepsilon^{acb}G^{c}_{\lambda\rho},
\ee
and from (\ref{comzero}) we will get the algebraic equation 
\be\label{commfield}
 \varepsilon^{acb}G^{c}_{\lambda\rho}  G^{b}_{\mu\nu} =0
\ee
or, in an equivalent form, that
\be
 [G_{\lambda\rho},  G_{\mu\nu}] =0.
\ee
It follows that the field strength tensor factorises into the product  of Lorentz tensor and colour vector, which is in the direction of the Cartan's sub-algebra.   Both fields can depend on the space-time coordinates:
\be\label{covconfac}
G^{a}_{\mu\nu}(x)= G_{\mu\nu}(x) n^a(x) .
\ee
Let us consider the following change of variables in the equation (\ref{YMeqcov}) \cite{Batalin:1976uv, Brown:1975bc, Duff:1975ue}
\be\label{subst}
A^{a}_{\mu} + {1\over 2} G^{a}_{\mu\nu}(A) x_{\nu} = a^a_{\mu}, 
\ee
so that by using the equation (\ref{commfield}) one can get
\be\label{defequ}
(\delta^{ab}\partial_{\rho} - g \varepsilon^{acb}a^{c}_{\rho}) G^b_{\mu\nu}  =0,~~~         \partial_{\mu} a^{a}_{\nu} - \partial_{\nu} a^{a}_{\mu}
 - g \varepsilon^{abc} a^{b}_{\mu} a^{c}_{\nu}=0.
\ee
It follows that the field $a_{\mu}$ is a flat connection 
\be\label{flatconn}
a_{\mu}  = - {i \over g} S^{-1} \partial_{\mu} S,
\ee
and the equation (\ref{subst}) will take the following form:
\be
A_{\mu} + {1\over 2} G_{\mu\nu}(A) x_{\nu}  = -  {i\over g} S^{-1} \partial_{\mu} S.
\ee
The general solution that would express  $A_{\mu}$ in terms of $S$ were not found in the past investigations. Only in the case  when $S= 1$ one can find from  (\ref{flatconn}) that $a_{\mu}=0$ and from (\ref{defequ})  that $ \partial_{\rho} G^{a}_{\mu\nu} =0 $, therefore 
\be\label{consfield}
A^{a}_{\mu} = - {1\over 2} F_{\mu\nu} x_{\nu}   n^a , 
\ee
where $F_{\mu\nu} $ and $n^a$ are space-time independent parameters.  

The problem that remained unsolved was to identify how large is the class of covariantly-constant gauge fields defined by the equation  (\ref{YMeqcov}).  We will demonstrate that the class of covariantly constant gauge fields is much larger than the fields defined by the equation (\ref{consfield}) and that in general these fields can have nontrivial space-time dependence.  A wider class of covariantly constant gauge fields can be obtained through the nontrivial space-time dependence of the unit vector $n^a(x)$ in (\ref{covconfac}).

Let us consider the Cho Ansatz  \cite{Cho:1979nv, Cho:1980nx, Cho:2010zzb} (see also \cite{tHooft:1974kcl}, formula (2.17), \cite{Corrigan:1975zxj}, formula (2.11), and \cite{Biran:1987ae}):
\be\label{choansatz}
A^{a}_{\mu} =  B_{\mu} n^{a}  +
{1\over g} \varepsilon^{abc} n^{b} \partial_{\mu}n^{c},
\ee
where $B_{\mu}(x)$  is the Abelian Lorentz vector and $n^{a}(x)$ is a space-time dependent colour unit vector:
\be
n^{a} n^{a} =1,~~~~~~n^{a} \partial_{\mu} n^{a} =0.
\ee
The unit vector $n^a(x)$ describes two independent field variables. One can get convinced that the covariant derivative of the colour unit vector in the background field (\ref{choansatz}) is equal to zero \cite{Cho:1979nv, Cho:1980nx}:
\be\label{unitcovzero}
\nabla^{ab}_{\mu} n^b= \partial_{\mu} n^{a} -g \varepsilon^{abc} A^{b}_{\mu}  n^{c}  =0,
\ee
and therefore 
\be
 [\nabla_{\mu} , \nabla_{\nu}]^{ab}  n^{b} = -g  \varepsilon^{acb}G^{c}_{\mu\nu} n^{b}  =0.
\ee
By calculating the field strength tensor (\ref{fieldstregth}) one can get convinced that the Yang-Mills  field strength tensor factorises  \cite{Cho:1979nv, Cho:1980nx}:
\be\label{chofact}
G^{a}_{\mu\nu} =  ( F_{\mu\nu} + {1\over g} S_{\mu\nu})~ n^{a},
\ee
where
\be\label{spacetimefields}
F_{\mu\nu}= \partial_{\mu} B_{\nu} - \partial_{\nu} B_{\mu},~~~~~~~~~
S_{\mu\nu}= \varepsilon^{abc} n^{a} \partial_{\mu} n^{b} \partial_{\nu} n^{c}.
\ee
There are two contributions to the field strength tensor $G^{a}_{\mu\nu}$, the first one is from the Abelian vector field $B_{\mu}(x)$ and the second one is from the colour unit vector $n^a(x)$. It is useful to parametrise the unit vector in terms of spherical angles \cite{Cho:1979nv, Cho:1980nx}:
\be
n^a = (\sin\theta \cos\phi, \sin\theta \sin\phi, \cos\theta ),
\ee
and then express $S_{\mu\nu}$ also in terms of spherical angles:
\be
S_{\mu\nu} = \sin\theta ( \partial_{\mu}  \theta  \partial_{\nu} \phi  - \partial_{\nu}  \theta   \partial_{\mu} \phi).
\ee
It follows then that $S_{\mu\nu}$  is a field strength tensor of the  colour field $C_{\mu}$:
\be\label{coulorvect}
S_{\mu\nu} =  \partial_{\mu}  C_{\nu} -   \partial_{\nu} C_{\mu} , ~~~~C_{\mu}= -\cos\theta  \partial_{\mu} \phi.
\ee
The colour field strength tensor $S_{\mu\nu}$ can be expressed either in terms of colour unit vector $n^a$, as in (\ref{spacetimefields}), or in terms of colour field $C_{\mu}$, as in (\ref{coulorvect}).  This fact is not accidental because the two-form $ d (S_{\mu\nu} dx_{\mu} \wedge d x_{\nu}) =0$    is an exact form and is therefore the derivative of one-form $C_{\mu} dx_{\mu}$.  

These   are the general properties of the Cho Ansatz,  and, as one can see, one of the important properties of the Cho Ansatz is that the field strength tensor factorises into the Lorentz and colour  structures  (\ref{chofact}). This factorisation is identical to the factorisation (\ref{covconfac}) that takes place in the case of covariantly constant gauge fields (\ref{YMeqcov}). It is therefore natural to search for the new solutions for covariantly constant gauge fields (\ref{YMeqcov}) in the form of  Cho Ansatz. We will formulate the conditions at which the Ansatz represents the covariantly constant gauge fields and is also a solution of the Yang-Mills equation.   In the case of Cho Ansatz  the Yang-Mills equation reduces to the following equation:
 \be
 n^{a} \partial_{\mu}  ( F_{\mu\nu} + {1\over g} S_{\mu\nu})=0,
 \ee
meaning that if $F_{\mu\nu} + {1\over g} S_{\mu\nu}$ is a solution of the Maxwell equations, then it can be also the solution of the Yang Mills equation.  We are interested in finding the conditions at which the Cho Ansatz represents the covariantly constant gauge field.  In that case the following equation should be fulfilled:
 \be
 n^a \partial_{\rho}  ( F_{\mu\nu} + {1\over g} S_{\mu\nu})=0,
 \ee
 meaning that the sum of terms in the brackets should be a constant tensor:
 \be\label{nakefield}
G_{\mu\nu}=  F_{\mu\nu} + {1\over g} S_{\mu\nu},  
 \ee
where $G_{\mu\nu} $ is an antisymmetric constant  tensor. We see that the moduli space of covariantly constant gauge fields increases because now we have two independent tensors to construct solutions:  the Abelian  tensor $ F_{\mu\nu}$ and the colour tensor $S_{\mu\nu}$.  Both of them can be space-time independent tensors or  the space-time dependence of  $ F_{\mu\nu}$ and $S_{\mu\nu}$ cancels in the sum.  Thus one can search for the solutions when both tensors are constant tensors or can be space-time dependent and this dependence cancels in the sum (\ref{nakefield}).  In other words, the moduli  space of the covariantly constant gauge fields can be much larger than the moduli  space of the solution (\ref{consfield}). The nontrivial solutions will be presented in the next sections.

 \section{\it Trigonometric Solutions } 
\begin{figure}
 \centering
\includegraphics[angle=0,width=6cm]{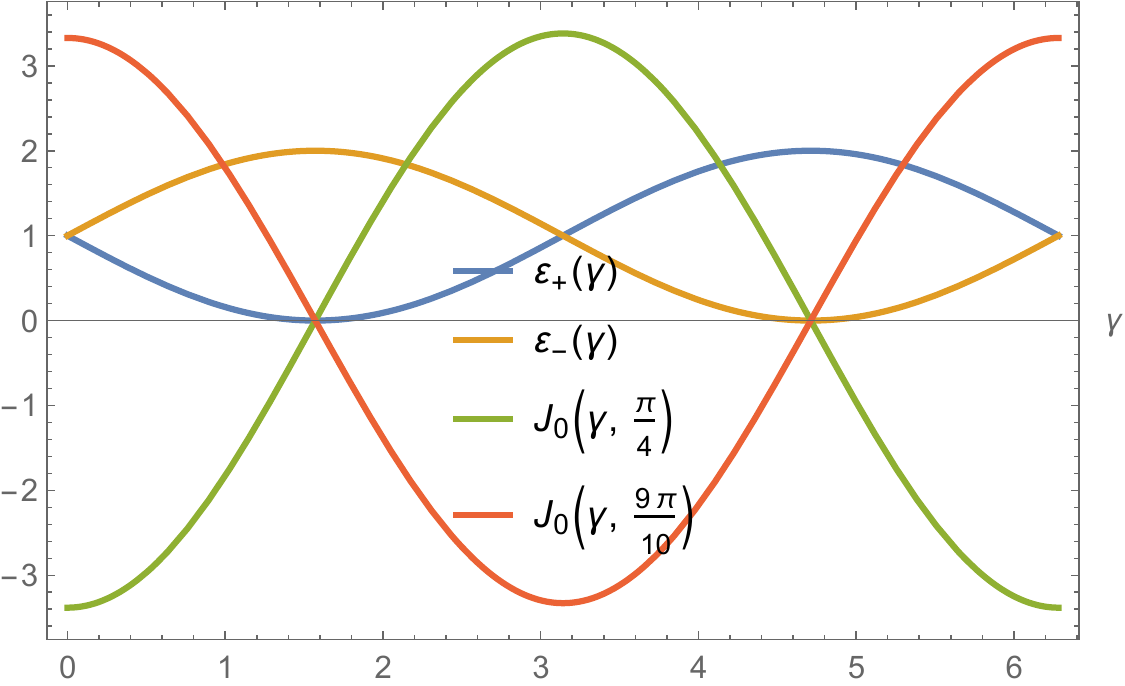} 
\includegraphics[angle=0,width=9cm]{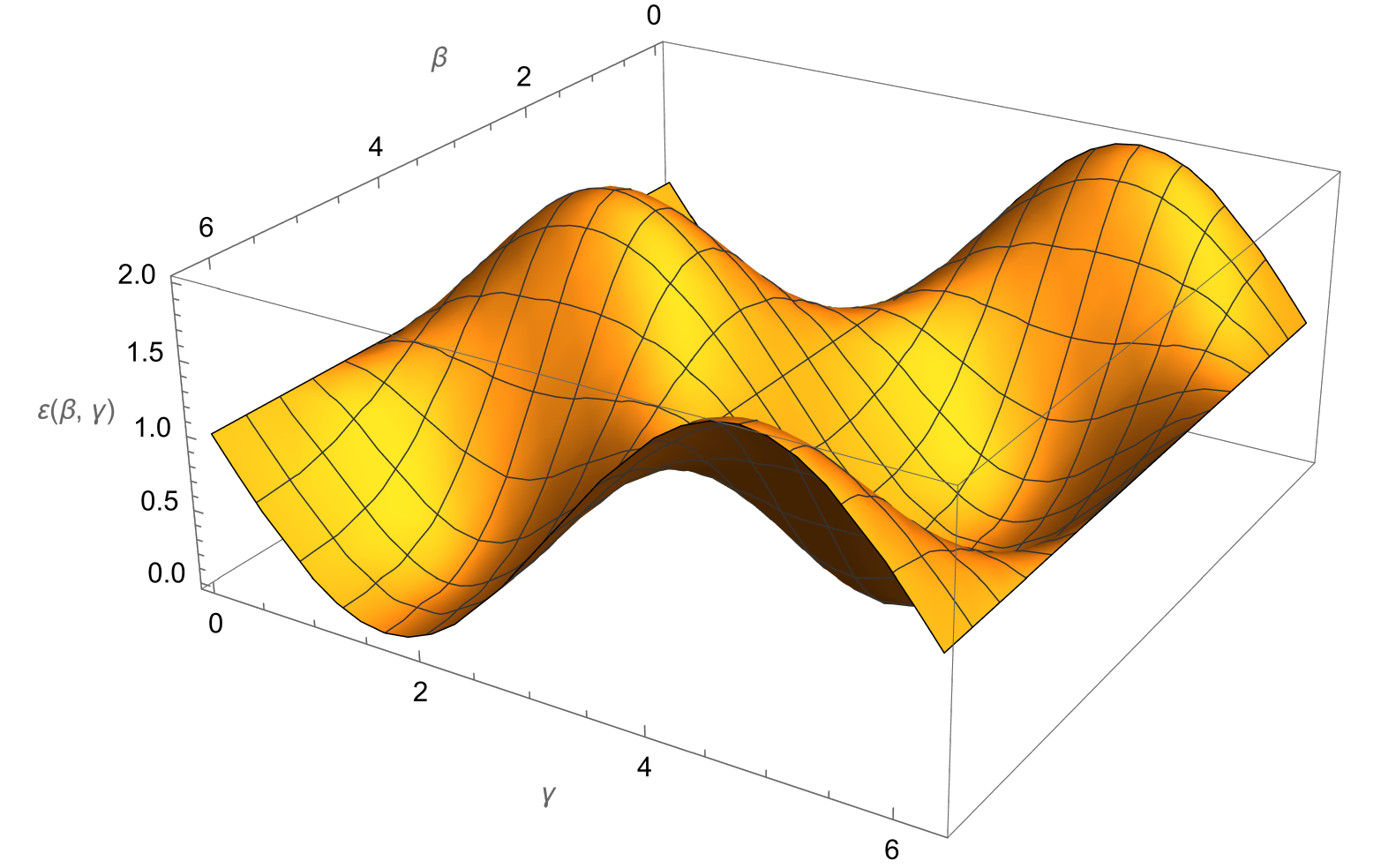} 
 \centering
\caption{The l.h.s. graph demonstrates the magnetic energy $\epsilon(\gamma)$ (\ref{density}) and the Hopf $J_0(\gamma,x)$ density  (\ref{Hopfdensity})  when the vectors are  $\vec{H}=(0,0,1)$, $\vec{a}=(1,0,0)$, $\vec{b}=(0,\cos\gamma,\sin\gamma)$. The density $J_0(\gamma,x)$ vanishes only when the vectors $\vec{H}$  and $(\vec{a}\times \vec{b})$ are parallel or antiparallel $(\gamma = \pi/2, 3\pi/2)$ and realise two minimums of the energy density.   For the field configuration $\vec{H}=(0,0,H)$, $\vec{a}=(a\cos\beta, a \sin\beta,0)$, $\vec{b}=(0,b \sin\gamma, b \cos\gamma)$ we will get the magnetic energy  landscape $\epsilon(\beta,\gamma)$ as a function of two angles  shown on the r.h.s. graph with a series of minimums separated by potential barriers. }
\label{fig2} 
\end{figure}

 We will consider the Abelian field $F_{\mu\nu}$ to be a constant tensor (\ref{spacetimefields})
\be\label{abelcovfield}
B_{\mu}= -{1\over 2} F_{\mu\nu} x_{\nu}
\ee
and will define our Ansatz for the unit colour vector in the following form:
\be\label{ansatz2}
n^a(\vec{x})= \{\sin(a\cdot x) \cos\Big({(b\cdot x) \over \sin(a\cdot x)}\Big),~\sin(a\cdot x) \sin\Big({(b\cdot x) \over \sin(a\cdot x)}\Big),~ \cos(a\cdot x)   \}, 
\ee 
 where $a_{\mu}$ and $b_{\nu}$ are constant vectors and $(a\cdot x) = a_{\mu} x_{\mu}$, $(b\cdot x) = b_{\mu} x_{\mu}$. The components of the   colour field $C_{\mu}$ defined in (\ref{coulorvect}) are 
\beqa\label{tridsol}
C_{\mu}= a_{\mu} (b\cdot x)   \cot^2(a\cdot x)  - b_{\mu} \cot(a\cdot x)   
\eeqa 
and  are singular on the planes  $a_{\mu} x_{\mu} = \pm \pi   n$, $n=0,1,2,....$. The corresponding field strength is a constant tensor
 \be
 S_{\mu\nu} =  \partial_{\mu}  C_{\nu} -   \partial_{\nu} C_{\mu} =  a_{\mu} \wedge b_{\nu} 
 \ee
and  is perfectly regular in the whole space. The space distance between neighbouring singularities is of order $1 /\vert \vec{a} \vert$ and tends to zero as $\vert \vec{a} \vert \rightarrow \infty$.  The  field strength tensor $G^a_{\mu\nu}$ has contribution from two vector fields: the Abelian $B_{\mu} $ and colour field  $C_{\mu}$ of the Ansatz (\ref{choansatz}) and has the following form:
 \be\label{consfielstr}
 G^{a}_{\mu\nu} = (F_{\mu\nu} + {1\over g} a_{\mu} \wedge b_{\nu} ) n^a(x).
 \ee
 The square of the field strength tensor is 
 \be\label{actioncont}
 {1\over 4 }G^{a}_{\mu\nu} G^{a}_{\mu\nu} = {1\over 4 } F_{\mu\nu} F_{\mu\nu} + {a_{\mu} F_{\mu\nu} b_{\nu} \over g}  +  { a^2 b^2 - (a\cdot b)^2  \over 2 g^2}.
 \ee  
 If the vectors $a_{\mu}$ and $b_{\nu}$ are parallel, then the contribution from the space-time dependent colour unit vector $n^a(x)$ will vanish and the field strength tensor reduces to the space-independent Abelian field $F_{\mu\nu}$.  The magnetic energy density can be represented in the following form:
\beqa\label{trigenergydens}
\epsilon &=& {\vec{H}^2 \over 2} - {1\over g} \vec{H} \cdot (\vec{a}\times \vec{b}) + {1\over 2 g^2} (\vec{a}\times \vec{b})^2 .
\eeqa  
The solution has a larger degeneracy because one can obtain different field configurations that have the same energy density  (\ref{trigenergydens}). 
The minimum of $\epsilon$ is realised when the term $ \vec{H} \cdot (\vec{a}\times \vec{b})$ gets its maximum positive value. This takes place when all three vectors  $ (\vec{H}, \vec{a}, \vec{b})$ are forming the right parallelepiped, so that 
$$\epsilon_{min} ={H^2 \over 2} - {1\over g} H a b + {1\over 2 g^2}  a^2 b^2 .$$   
Let us consider the field configuration  $\vec{H}=(0,0,H)$, $\vec{a}=(a\cos\beta, a \sin\beta,0)$, $\vec{b}=(0,b \sin\gamma, b \cos\gamma)$  parametrised by two angles $\beta$ and $\gamma$, so that the magnetic energy  landscape $\epsilon(\beta,\gamma)$ as a function of two angles has a series of minimums separated by potential barriers (see Fig.\ref{fig2}):
\be\label{density}
\epsilon(\beta,\gamma)=     {H^2 \over 2} - {1\over g} H a b \sin\gamma \cos\beta +  { a^2 b^2 \over 2  g^2}(\cos\gamma ^2 + \sin\gamma^2 \cos\beta^2 ).
 \ee
Here we have three independent vectors  $(\vec{H},\vec{a},\vec{b})$.  Geometrically the magnetic "force lines" of the Abelian field $B_{\mu}$ are along the $z$ axis and the magnetic force lines of the colour  field $C_{\mu}$ are in the direction $(\vec{a} \times \vec{b})$, so that they cross each other under the angle $\gamma$.  As we will see in the next section, this geometrical configuration of the magnetic fields has a nontrivial Hopf invariant density  when the vectors $\vec{H}$ and $(\vec{a}\times \vec{b})$ are not parallel.

 The appearance of singular planes in the gauge potential (\ref{tridsol}) hints to the fact that here we may have a nontrivial monopole density distributed over the planes.  As we will demonstrate shortly, the discontinuities at the singular planes $a_{\mu} x_{\mu} = \pm \pi   n$, $n=0,1,2,....$ can be related by a gauge transformation, similar to the discontinuity of the monopole field on the equator where two gauge field patches intersect   \cite{Wu:1975es, Wu:1975vq}. In the next section we will define the topological currents and the corresponding  charges  and will demonstrate that the monopole magnetic density vanishes for our solution. Instead, the solutions have a nonzero Hopf invariant density distributed over the whole space. 
 
 \section{\it Topological Currents and  Hopf Helicity Invariant }
 
The conserved  topological  current and the corresponding monopole charge is defined in terms of the field strength $G_{\mu\nu}$ (\ref{nakefield}) and the gauge field $A_{\mu}=B_{\mu} +  {1\over g } C_{\mu}$    \cite{Cho:1979nv, Cho:1980nx}:
 \be\label{chocurent}
K_{\mu} = \epsilon_{\mu\nu\lambda\rho} \partial_{\nu} G_{\lambda\rho}   =\epsilon_{\mu\nu\lambda\rho} \partial_{\nu} ( F_{\lambda \rho} + {1\over g} S_{\lambda\rho})  
,  ~~~~~~q_m= \int K_{0}d^3 x . 
 \ee
 The background for this proposal is based on the fact that the Yang-Wu singular monopole solution \cite{Wu:1975vq, Wu:1967vp}  can be obtained   when $n^a = {x^a \over r}$  ($B_{\mu}= F_{\mu\nu}=0$)   \cite{Cho:1979nv, Cho:1980nx}:
\be
 C^a_{i}=   {1\over g}\varepsilon^{abc} n^{b} \partial_{i}n^{c}= \epsilon^{aij}{x^j\over r^2},~~~~~~S_{ij}= \varepsilon^{abc} n^{a} \partial_{i} n^{b} \partial_{j} n^{c}= \epsilon_{ijk} {x^k \over r^3},~~~~q_m ={1\over g} \int \varepsilon_{ijk} \partial_{i} S_{jk}d^3 x,
\ee
and it has a topological degree,  the number of times a mapping of the space boundary covers the unit sphere $\CS^2_{space~boundary} \rightarrow \CS^2_{iso-sphere}$,  as well as  it is equal to the  total magnetic flux through the closed surface,  that is,  the monopole magnetic charge $q_m = 4\pi /g$ \cite{tHooft:1974kcl,Goddard:1976qe}. 

Our solution (\ref{tridsol}) has the constant field  strength (\ref{consfielstr})
 \be
  G_{\mu\nu}=F_{\mu\nu} + {1\over g} a_{\mu} \wedge b_{\nu},
 \ee
and  the topological current (\ref{chocurent}) vanishes, $K_{\mu}=0$, together with the monopole charge. Thus the solution has zero monopole density and therefore does not  describe the condensation of monopoles despite the existence of singular planes in the gauge potential. We don't  know if there exist covariantly-constant gauge fields that have a nonzero monopole charge density.  

Instead, we would define the topological current and the corresponding Hopf invariant $h$ \cite{Hopf:1931,Whitehead:1947},  which measures the helicity of a divergent-free magnetic field and is equal to the linking of the Faraday "lines of force":
 \be\label{savtopcharge}
J_{\mu} = \epsilon_{\mu\nu\lambda\rho} G_{\nu\lambda} A_{\rho}   =\epsilon_{\mu\nu\lambda\rho} ( F_{\nu\lambda} + {1\over g} S_{\nu\lambda}) (B_{\rho}+ {1\over g } C_{\rho} ),  ~~~~~~h= \int J_{0}d^3 x = \int \epsilon^{ijk} G_{ij} A_k . 
 \ee
The Hopf invariant $h$ is nonzero if the lines of force within a magnetic field are  twisted, knotted, or entangled\footnote{It is the abelian Chern-Simon action \cite{cmp/1104178138} for a gauge potential one-form $A$ on the 3d-space $S_{CS} = \int_{\CR^3} A \wedge dA$.} \cite{Moffatt:1969,Moffatt:1991}. 
The Hopf invariant density is non-vanishing for the covariantly constant gauge fields. The Hopf invariant  essentially differs from the monopole charge defined above and has an alternative geometrical meaning  \cite{Hopf:1931,Whitehead:1947,cmp/1104178138,Antoniadis:2012ep}, which will be discussed shortly.  We can calculate  the topological current defined in (\ref{savtopcharge}) in the following way:
 \beqa
J_{\mu}    = {1\over g } \epsilon_{\mu\nu\lambda\rho}  F_{\nu\lambda}   a_{\rho} (b\cdot x)   \cot^2(a\cdot x)  
-  {1\over g } \epsilon_{\mu\nu\lambda\rho}  F_{\nu\lambda}  b_{\rho} \cot(a\cdot x)   -{1\over  g} \epsilon_{\mu\nu\lambda\rho}   a_{\nu}  b_{\lambda}    F_{\rho\sigma} x_{\sigma}   ~~~~~
 \eeqa
 and the Hopf invariant density will take the following form:
 \be
 J_0 = {1\over g}(\vec{H}\cdot \vec{a})\Big( 2 (\vec{b}\cdot \vec{r}) \cot^2(\vec{a}\cdot \vec{r})  +(\vec{b}\cdot \vec{r})\Big) -  {1\over g}(\vec{H}\cdot \vec{b}) \Big( 2 \cot(\vec{a}\cdot \vec{r})  +(\vec{a}\cdot \vec{r})\Big).
 \ee
This density is non-zero  when the vectors $\vec{H}$ and $(\vec{a}\times \vec{b})$ are not parallel. In order to get convinced that our solution has a nonzero helicity density let us consider the configuration   $\vec{H}=(0,0,H)$, $\vec{a}=(a\cos\beta, a \sin\beta,0)$, $\vec{b}=(0,b \sin\gamma, b \cos\gamma)$,  so that  we will get
\beqa\label{Hopfdensity}
J_0 =    {1\over g }H b \cos\gamma \Big( 2 \cot(a x \cos\beta + a y \sin\beta) + (a x \cos\beta + a y \sin\beta)\Big) .
\eeqa 
The vectors   $\vec{H}$ and $(\vec{a}\times \vec{b})$ define two systems of magnetic force lines.  The density $J_0$ vanishes only if these vectors are parallel or antiparallel $(\gamma = \pi/2, 3\pi/2)$. Considering the projection (compactification) of the space $\CR^3$ into the 3-sphere $\CS^3$ one can observe that these magnetic force lines are winding around each other and therefore  the Hopf invariant is nonzero\footnote{In the case of the compact space the Hopf invariant distinguishes  the continuous  mappings from $\CS^3$ to $\CS^2$ and is equal to the linking number of two curves $(\gamma_1,\gamma_2)$ that are preimages of two arbitrary points of the two-sphere.  It has a topological meaning as the linking number and as the integral of the volume element on $\CS^3$. The Gauss'  linking integral is  defined as $lk(\gamma_1,\gamma_2)= {1\over 4\pi} \int_{\gamma_1} dx^i  \int_{\gamma_2} dx^j   \epsilon_{ijk} {(x-y)^{k}\over \vert x-y\vert^3 }$.} . 
Thus our solution has a non-vanishing  Hopf invariant  density distributed over the whole space. 

The conclusion is that the solution has a zero monopole charge density and therefore  does not  describe the condensation of monopoles despite the existence of singular planes in the gauge potential.   We don't  know if solutions with nonzero magnetic monopole density exist within the covariantly constant gauge fields.  Non-vanishing of the Hopf density means that the solution cannot be continuously deformed to a constant chromomagnetic field (\ref{consfield}) and we have a degeneracy of  the classical vacuum.  This degeneracy is in addition to the degeneracy due to the nontrivial flat connections $A_i =S^{-1} \partial_{i} S$, where $S$ is the unitary matrix of a gauge transformation that cannot be joined to the identity through the continuous  transformations \cite{Jackiw:1976pf,Callan:1976je}.    In the next section we will present additional solutions of the covariantly constant field equation.  
 
\section{\it Chromomagnetic Flux Sheets }

We found a "chromomagnetic flux sheet"  solution of the Yang-Mills equation when the unit vector field has the following form:
\be\label{polsol}
n^a(x)= \{ \sqrt{1-(a\cdot x)^2} \cos(b\cdot x),~ \sqrt{1-(a\cdot x)^2} \sin(b\cdot x),~ -(a\cdot x)  \},
\ee 
 where $a_{\mu}$ and $b_{\nu}$ are arbitrary constant Lorentz vectors  and $(a\cdot x) = a_{\mu} x_{\mu}$. We can calculate the colour field $C_{\mu}$ and the  corresponding field strength tensor $S_{\mu\nu}$ defined in (\ref{spacetimefields}) and (\ref{coulorvect}):
 \be\label{cfield2}
 C_{\mu}= b_{\mu}  (a\cdot x), ~~~~   S_{\mu\nu} = a_{\mu} b_{\nu} - a_{\nu} b_{\mu} \equiv   a_{\mu} \wedge b_{\nu}, 
 \ee
 so that  the full field strength tensor is equal to the following expression:
 \be
 G^{a}_{\mu\nu} = (F_{\mu\nu} + {1\over g} a_{\mu} \wedge b_{\nu} ) n^a(x) ,
 \ee
 where the tensor $F_{\mu\nu}$ is a constant tensor. The square of the field strength tensor is 
 \be\label{actioncont1}
 {1\over 4 } G^{a}_{\mu\nu} G^{a}_{\mu\nu} = {1\over 4 } F_{\mu\nu} F_{\mu\nu} + {a_{\mu} F_{\mu\nu} b_{\nu} \over g}  +  {a^2 b^2 - (a\cdot b)^2 \over 2 g^2}.
 \ee
 \begin{figure}
 \centering
\includegraphics[angle=0,width=5cm]{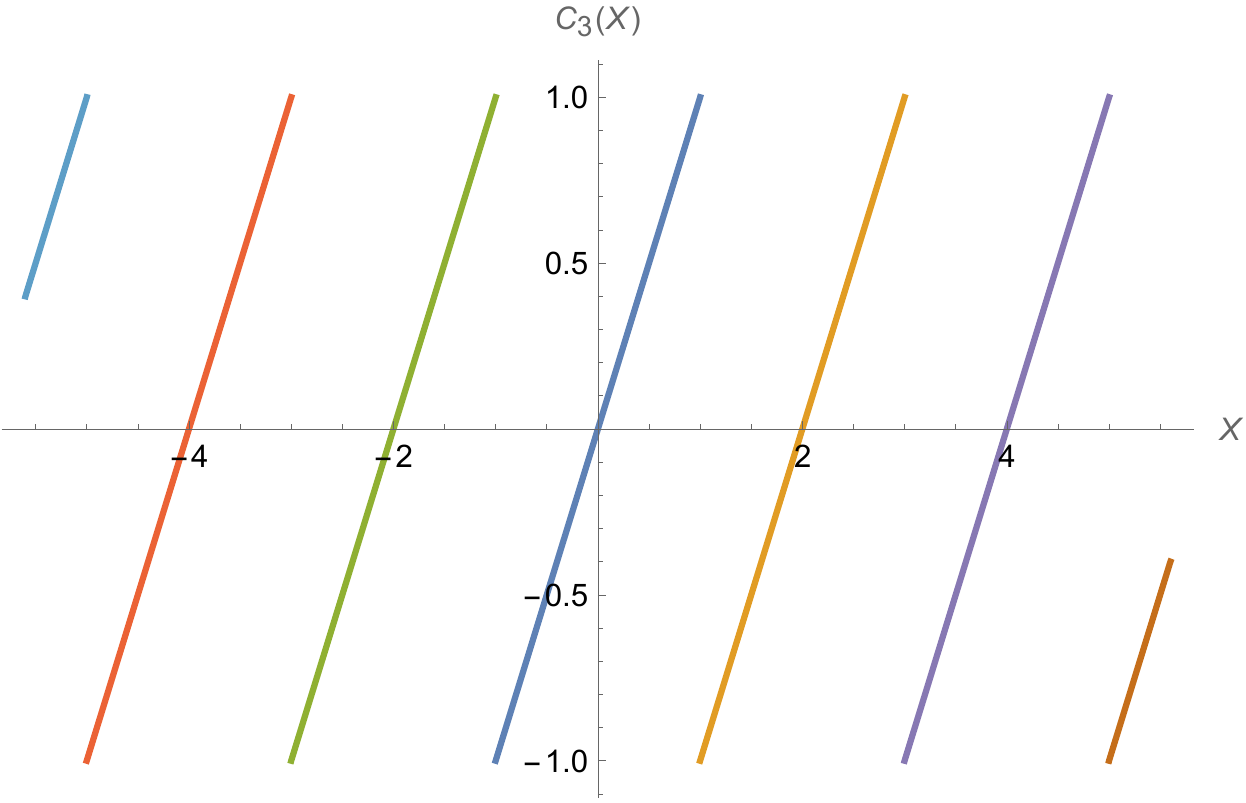}~~~~~~
\includegraphics[angle=0,width=5cm]{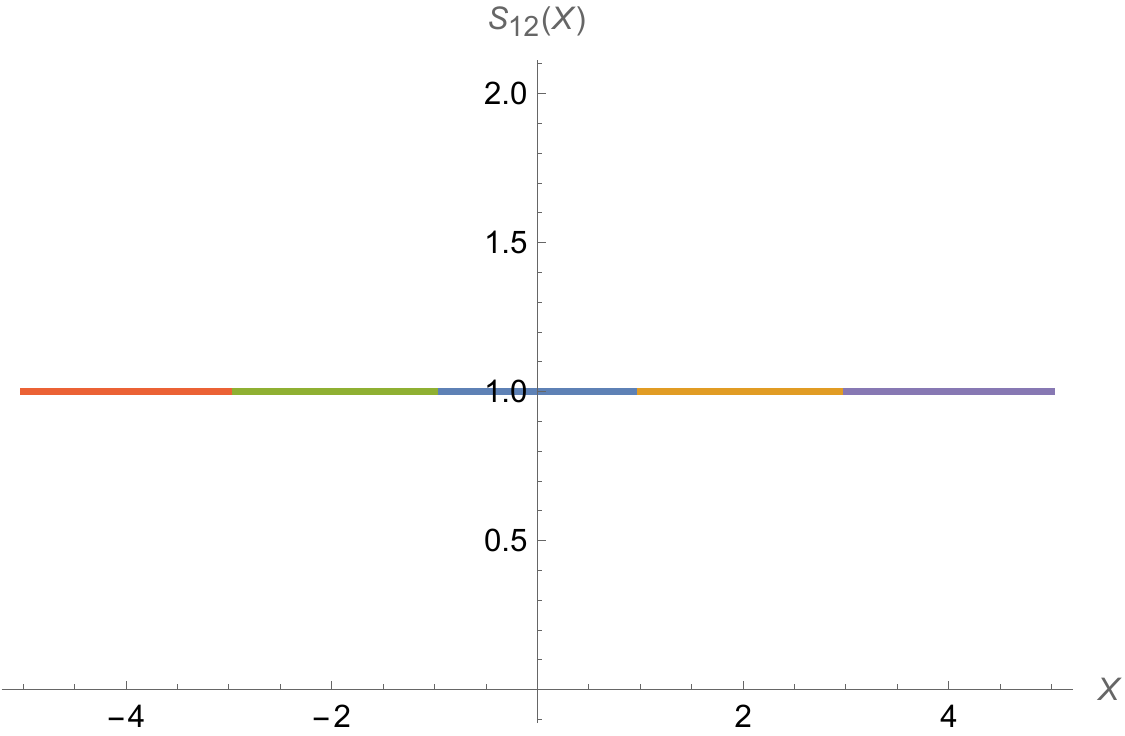}
\centering
\caption{The graph demonstrates the space variation of the $C_3(x)$  component of the  colour field $C_{\mu}$ (\ref{cfield}) and of the corresponding field strength tensor $S_{13}$ (\ref{cfield1}) when $\vec{a}=(1,0,0)$, $\vec{b}=(0,0,1)$.  The coordinates $y,z$ are not shown. The discontinuities of the $C_3(x)$  are at the planes $x=\pm n, n=0,1,2...$, and the derivatives of the $C_3(x)$ from the l.h.s and from the r.h.s  are equal, and therefore the  field strength tensor $S_{13}$ is perfectly regular everywhere.  The discontinuities $C_3(n-\epsilon)-C_3(n+\epsilon)=\partial_z \Phi$ are related by the gauge transformation  $\Phi =2 z$, similar to the discontinuity of the monopole field on the equator where two gauge field  patches intersect   \cite{Wu:1975es, Wu:1975vq}.  The action density (\ref{actioncont1}) is a space constant, and the topological  density (\ref{savtopcharge}) is $J_0 =  F_{12} a_1 b_3 x=  H x $.}
\label{fig1} 
\end{figure}
The magnetic energy density can be represented in the following form: 
\beqa
\epsilon &=& {\vec{H}^2 \over 2} - {1\over g} \vec{H} \cdot (\vec{a}\times \vec{b}) + {1\over 2 g^2} (\vec{a}\times \vec{b})^2 .
\eeqa
One can verify explicitly that the gauge field (\ref{choansatz}), (\ref{abelcovfield}) and (\ref{polsol})  is a solution of the Yang-Mills equation and is defined on a sheet:
 \be\label{wall}
 (a\cdot x)^2 \leq 1.  
 \ee 
Outside of the sheet $ (a\cdot x)^2 > 1 $ all fields are defined to vanish.  Let us consider the solution when $B_{\mu}=F_{\mu\nu}=0$, so that 
\beqa\label{polsol1}
A^{a}_{\mu} =   {1\over g} \varepsilon^{abc} n^{b} \partial_{\mu}n^{c},~~~~~~~~G^{a}_{\mu\nu} =  {1\over g} (a_{\mu} \wedge b_{\nu} ) n^a(x), ~~~~~~~~~~~
\epsilon =  {1\over 2 g^2} (\vec{a}\times \vec{b})^2 .
\eeqa
The solution (\ref{polsol}),  (\ref{polsol1})  is defined on a 3D-space sheet of a finite thickness, and we  defined the field $n^a(x)$ outside of the sheet to vanish  (see Fig.\ref{fig3}).  In that case  the solution represents a {\it non-perturbative  magnetic sheet of finite thickness $\propto 1/\vert a \vert $}. For simplicity  let us consider the parameters $a_{\mu}=(0,a,0,0)$ and $b_{\nu}=(0,0,0,b)$.  In that case the gauge field has the following components 
\beqa\label{magneticsheetsolution}
A^{a}_{\mu} &=&   {1\over g} \left\{
\begin{array}{cccc}   
(0,0,0)& \\
\Big({a \cos b z\over \sqrt{1-(a x)^2}}, -{a \sin b z\over \sqrt{1-(a x)^2}}, 0\Big)&  (a  x)^2 \leq 1\\
(0,0,0)&\\
\sqrt{1-(a x)^2} \Big(a  b x \sin b z , a  b x \cos b z ,-b \sqrt{1-(a x)^2}~\Big)
\end{array} \right.  \\
A^{a}_{\mu} &=&  0,~~~~~~~~~~~~~~~~~~~~~~~~~~~~~~~~~~~~~~~~~~~~~~~~~~~~~~~~~~~~~~~~~~~~~~~ (a  x)^2 \geq 1.\nn
\eeqa 
One can verify explicitly  that it is a solution the Yang Mills equation. The  field strength outside of the magnetic sheet vanishes. There is no energy flow from the magnetic sheet in the direction transversal to the sheet  because the Poynting vector vanishes, $\vec{E^a} \times \vec{H^a} =0$.  In some sense these solutions are similar to the Nielsen-Olesen magnetic flux tubes, and are supported without presence of any Higgs field. The quantum mechanical stability of the solution remains to be investigated. 

One  can consider the limit $a \rightarrow \infty $ of infinitely thin surface on which we will have a flux of chromomagnetic field. In the opposite limit $a \rightarrow 0 $ the flux will spread all over the 3D-space.  One can also construct  the solution that describes a finite or infinite many parallel layers of flat magnetic sheets distributed over the 3D-space.   It seems that this singular surface solutions can be associated  with the  singular surfaces  considered by 't Hooft in \cite{tHooft:1981bkw} where he discussed  a possible existence of such non-perturbative solutions (see also \cite{Kapustin:2005py, Kapustin:2006pk, Gukov:2006jk, Gukov:2008sn}).  

 Let us extend this solution to the whole space-time keeping the parameter $a$ fixed.  For transparency let us consider a particular vector $a_{\mu}= (0, 1, 0, 0)$ in $x$ direction and define the field in the region $1 \leq x \leq 3$ as   $$n^a(x)= \{ \sqrt{1 -( x-2)^2} \sin(b\cdot x),~ \sqrt{1-(x-2)^2} \cos(b\cdot x),~  -(x - 2) \}$$  so that the field strength is  a continuous function at the point $x=1$. One can  extend the vector field into the whole line and define  the corresponding gauge field $C_{\mu}$ as 
 \beqa\label{cfield}
C_{\mu}= -\cos \theta  \partial_{\mu} \phi = -\cos \theta b_{\mu},~~~~
-\cos \theta  =\left\{
  \begin{array}{ccccccc}
      ...    & ...   \\
      x+2 & ~~-3 \leq  x  \leq -1\\
      x    &~~  -1 \leq  x  \leq  1 \\
      x-2 &~~~~~ 1 \leq  x  \leq  3\\
      ... & ...
   \end{array} \right. 
\eeqa
The discontinuities $C_{\mu}(n-\epsilon)-C_{\mu}(n+\epsilon)=\partial_{\mu} \Phi$ are related by the gauge transformation  $\Phi =2 (b\cdot x)$, similar to the discontinuity of the monopole field on the equator where two patches intersect  \cite{Wu:1975es, Wu:1975vq}.  
In the whole space-time the corresponding field strength tensor has the following form (see Fig.\ref{fig1}): 
\be\label{cfield1}
S_{\mu\nu} = \delta_{\mu 1} b_{\nu}  -  \delta_{\nu 1} b_{\mu}.
\ee
 As one can see, the chromomagnetic field is a regular function in the whole space (see Fig.\ref{fig1}) and the Hopf density is 
 \be\label{savtopcharge3}
 J_0 = {1\over g } \epsilon_{ijk}(F_{ij} b_k (a\cdot x) - a_i b_j F_{kl}x_l)=  {1\over g } \Big(( \vec{a}\cdot \vec{H})( \vec{b}\cdot \vec{r})+ ( \vec{b}\cdot \vec{H})( \vec{a}\cdot \vec{r})\Big).
 \ee
 \begin{figure}
 \centering
\includegraphics[angle=90,width=4cm]{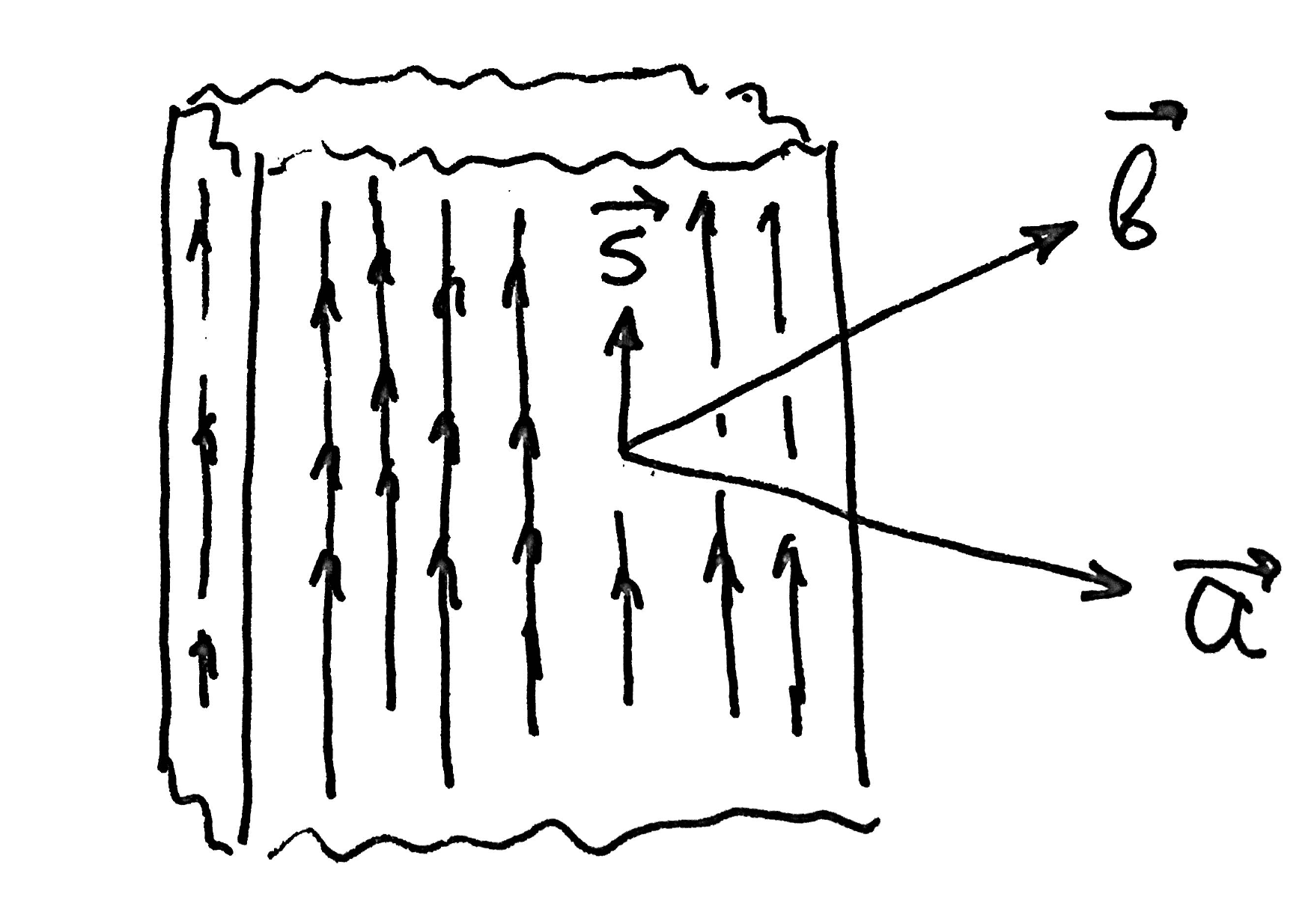}
\centering
\caption{The figure demonstrates a sheet of  finite thickness in the 3D-space that is filled by a  chromomagnetic field.  It represents a non-perturbative  magnetic sheet solution  (\ref{polsol}),  (\ref{polsol1}). The chromomagnetic  field $s_i = {1\over 2} \epsilon_{ijk} S_{ij}$  is nonzero inside the 3D-space sheet (\ref{cfield2}) and its direction is defined by the cross product $\vec{s} = (\vec{a} \times \vec{b})$. It vanishes outside the sheet.   
 }
\label{fig3} 
\end{figure}
For the field configuration $\vec{H}=(0,0,H)$, $\vec{a}=(a\cos\beta, a \sin\beta,0)$, $\vec{b}=(0,b \sin\gamma, b \cos\gamma)$ we will get the magnetic energy  landscape $\epsilon(\beta,\gamma)$ as a function of two angles with a series of minimums separated by potential barriers (see Fig.\ref{fig2}):
 \be
 \epsilon=     {H^2 \over 2} - {1\over g} H a b \sin\gamma \cos\beta +  { a^2 b^2 \over 2  g^2}(\cos\gamma ^2 + \sin\gamma^2 \cos\beta^2 ),
 \ee
and the Hopf invariant density   (\ref{savtopcharge3}) is 
\be
J_0 =    {1\over g }H a b \cos\gamma   (x  \cos\beta +   y \sin\beta). 
\ee

\section{\it General Solution}

One can guess that the moduli space of covariantly constant gauge fields is much larger  and can be obtained by solving the following system of partial differential equations:
\beqa\label{gen}
S_{12}= \sin \theta (\partial_1 \theta \partial_2 \phi - \partial_2 \theta \partial_1 \phi ) \nn\\
S_{23}= \sin \theta (\partial_2 \theta \partial_3 \phi - \partial_3 \theta \partial_2 \phi ) \nn\\
S_{13}= \sin \theta (\partial_1 \theta \partial_3 \phi - \partial_3 \theta \partial_1 \phi ),
\eeqa
where $S_{ij}$ are constants. The linear combination of these equations defines the  angle $\phi$ as an arbitrary function of  the variable 
$Y=  b_1 x +b_2 y + b_3 z $,
thus
\be
\phi = \phi(X) = \phi( b_1 x +b_2 y + b_3 z )=\phi( b\cdot x ).
\ee
After substituting the above function into the equations (\ref{gen}) one can observe that the angle $\theta$ is an arbitrary function of the alternative variable $X= a_1 x +a_2 y + a_3 z$, thus
\be
\theta=\theta(Y)=  \theta(a_1 x +a_2 y + a_3 z)=\theta( a\cdot x ).
\ee
It follows then that  three equations (\ref{gen})  reduce to the following differential  equations:
\be\label{genecovfie}
S_{ij} = a_i \wedge b_j  \sin\theta(X) ~ \theta(X)^{'}_X   ~ \phi(Y)^{'}_Y , 
\ee
where the derivatives are over the respective arguments.  If we are interested in finding the solutions with a constant tensor $S_{ij}$,  then the following condition should be fulfilled:
\be\label{ansatz7}
\sin\theta(X) ~ \theta(X)^{'}_X   ~ \phi(Y)^{'}_Y =1,
\ee
so that 
\be
S_{ij} = a_i \wedge b_j  .
\ee
The variables in (\ref{genecovfie})  are independent, and one can choose an arbitrary function for the angle $\theta$ and define the angle $\phi$. In particular, if $\theta(X)^{'}_X =1$, we will have $\theta(X) = X + X_0$ and $\phi = Y/\sin(X +X_0)$, thus recovering the solution (\ref{ansatz2}). If $\theta(X) $ is an arbitrary function of $f(X)$, then $\phi = Y/\sin f(X) f(X)^{'}_X$, and we have the following solution for the colour unit vector:
\be\label{ansatz5}
n^a(\vec{x})= \{\sin f( X)  \cos\Big({Y \over f^{'}(X) \sin(f( X))}\Big),~\sin(f( X)) \sin\Big({Y \over f^{'}(X)  \sin(f( X))}\Big),~ \cos(f( X))   \}.
\ee 
In the particular case when $f(X)=f(x)$,  it will reduce to the solution
 \be\label{ansatz3}
n^a(\vec{x})= \{\sin f( x)  \cos\Big({(b\cdot x) \over f^{'}(x) \sin(f( x))}\Big),~\sin(f( x)) \sin\Big({(b\cdot x) \over f^{'}(x)  \sin(f( x))}\Big),~ \cos(f( x))   \},
\ee 
where $f(x)$ is an arbitrary function of the space coordinate $x$ and $b_{\mu}=(b_0,b_1,b_2,b_3)$. The expressions for $C_{\mu}$ and $S_{\mu\nu}$ are too large to be presented but the square of the fields strength tensor is compact and has the following form:
\be
\epsilon =   {\vec{H}^2 \over 2 }  + {H_2 b_3 - H_3 b_2\over g}   +  {b^2_2 +b^2_3 \over 2 g^2}.
\ee
The other solution is given by the formula 
\beqa\label{ansatz4}
n^a(\vec{x})= \{\sin f( a_1 x+a_2 y)  \cos\Big({z \over f^{'}(a_1 x+a_2 y) \sin f( a_1 x+a_2 y)}\Big),~\nn\\
\sin f( a_1 x+a_2 y)  \sin\Big({z \over f^{'}(a_1 x+a_2 y)  \sin f( a_1 x+a_2 y)}\Big),~ \cos f(a_1 x+a_2 y)   \},
\eeqa 
and the corresponding energy density has the following form:
\be
\epsilon =   {\vec{H}^2 \over 2 }  + {H_1 a_2 - H_2 a_1 \over g} +  {a^2_1 +a^2_2 \over  2 g^2}.
\ee
Thus the equation (\ref{ansatz7}) allows to define a large class  of solutions by choosing an arbitrary function $\theta$ and then obtaining $\phi$  by integration or, alternatively, by considering an arbitrary function for $\phi$ and obtaining $\theta$  by integration.  We conclude that the moduli space of covariantly constant gauge fields is indeed larger than (\ref{abelcovfield}).  In comparison, the moduli space $\CI_{k,N}$  of the YM self-duality equation in the Euclidean space has the dimension $dim \CI_{k,N}=4 k N$ for the $SU(N)$ group.  It would be interesting to describe precisely  the moduli space of the covariantly constant gauge fields  defined by (\ref{YMeqcov}) and (\ref{genecovfie}).
 
 \section{\it Conclusion } 
  
In the recent articles \cite{Savvidy:2022jcr, Savvidy:2023kmx, Savvidy:2023kft} we examined the phenomena of the chromomagnetic gluon condensation in the Yang-Mills theory and the problem of stability of the chromomagnetic vacuum fields (\ref{abelcovfield}).  It was demonstrated that the apparent instability of these chromomagnetic vacuum fields is a result of the quadratic approximation. The stability is restored when the nonlinear interaction of negative/unstable modes is taken into account in the case of chromomagnetic vacuum fields and the interaction of the Leutwyler zero modes in the case of (anti)self-dual covariantly constant vacuum fields.  The vacuum fields (\ref{abelcovfield}) appear to be stable and indicate that the Yang-Mills vacuum is a highly degenerate quantum state. This consideration was focused on the stability of the vacuum fields represented by the Abelian solution (\ref{abelcovfield}).   

Here we investigated the moduli space of covariantly constant  gauge field solutions representing the Yang-Mills classical vacuum.   The identification of the moduli space  of covariantly constant gauge fields  defined by the equation  (\ref{YMeqcov}) remained unsolved for a long time.  We found that this moduli space is much larger than previously expected (\ref{abelcovfield}) and can be described in terms of the Cho Ansatz.   
 
The existence of an even larger class of  covariantly constant gauge fields described above once again pointed out to the fact that the Yang-Mills vacuum is a highly degenerate  state. It is a challenging problem to investigate the vacuum polarisation induced by this new class of covariantly constant gauge fields. Each covariantly constant gauge field configuration on its own contains a rich diversity of an emergent nonperturbative structure. It is the process of averaging over these vacuum  field configurations  that restores the Lorentz invariance of the vacuum state. One can conjecture that the effective Lagrangian for all covariantly constant  gauge fields has a universal form  (\ref{YMeffective}).  

The monopole magnetic density of our solutions  vanishes, but they have a nontrivial Hopf topological density distributed over the whole space.  The non-vanishing of Hopf density means that the solution cannot be continuously deformed to a constant chromomagnetic field (\ref{consfield}), and we have a degeneracy of  the classical vacuum.   In some sense these solutions (\ref{magneticsheetsolution}) are similar to the Nielsen-Olesen magnetic flux tubes,  but instead have two-dimensional magnetic sheet geometry and are supported without presence of any Higgs field.  The infinitesimally thin sheet solutions can be associated  with the  singular surfaces considered by 't Hooft in \cite{tHooft:1981bkw} where he discussed  a possible existence of such non-perturbative solutions. The nonlocal operators that are supported on a two-dimensional surface rather than a one-dimensional curve where considered in \cite{Kapustin:2005py, Kapustin:2006pk, Gukov:2006jk, Gukov:2008sn}. These surface operators are analogous to Wilson $W(C)$ and ’t Hooft line operators $M(C)$ except that they are supported on a two-dimensional surface rather than a one-dimensional curve.

Do there exist physical systems that have high degeneracy of the vacuum state? The Yang-Mills theory in the classical approximation has an infinite degeneracy of the vacuum state, which is labelled by the topological winding number, the Pontryagin index, which defines several topologically inequivalent sectors separated by potential barriers. In the quantum theory the tunneling occurs across these barriers  and the classical degeneracy is lifted due to the instantons  \cite{Jackiw:1976pf,Callan:1976je}.  The degeneracy of the classical vacuum  due to the covariantly constant field appears in addition to the degeneracy due to the nontrivial flat connections $A_i =S^{-1} \partial_{i} S$, where $S$ is the unitary matrix of a gauge transformation that cannot be joined with the identity through continuous  transformations.    
 
The sourceless solutions  of  the Yang-Mills equation were  recently investigated in a series of articles \cite{Pak:2020izt, Pak:2020obo, Pak:2017skw, Pak:2020fkt}. These solutions represent time-dependent standing waves reminiscent to the homogeneous plane wave solutions \cite{Kim:2016xdn}. The other class of solutions of the Yang-Mills equation in the background field  was investigated in a series of articles  and has a  "spaghetti" type structure of magnetic tubes \cite{Nielsen:1978tr,Nielsen:1979xu, Ambjorn:1979xi,Ambjorn:1980ms}. It would be interesting to find the "average" value of the Hopf invariant in that case.

Turning to the statistical spin systems, one can observe that the classical 3D Ising spin system has a double degeneracy of the vacuum state $(\vert \Uparrow \rangle, \vert \Downarrow \rangle)$. It is this symmetry that allows to construct a dual gauge invariant representation  of the 3D Ising model \cite{Wegner:1971app}. The spin system that has {\it an exponential degeneracy of the vacuum state}  was constructed in \cite{Savvidy:1993ej}  and is an extension of the 3D Ising model to the system with {\it direct ferromagnetic  and  one quarter  nearest-to-neighbour  antiferromagnetic interactions}.  In that system the parallel planes of spins on which the spins are oriented in any of the two directions represent a vacuum configuration $(\vert \Uparrow \rangle, \vert \Downarrow \rangle, ...\vert \Uparrow \rangle, ....\vert \Downarrow \rangle, \vert \Uparrow \rangle)$. The total number of such vacuum configurations is $3 \times 2^{N}$ \cite{Savvidy:2000zq}.  There is intriguing  similarity between the geometry of {\it magnetic sheets of spins} in the spin system and {\it the solutions representing magnetic sheet configurations of a finite width in the Yang Mills theory} (\ref{cfield2}) and (\ref{wall}). The spin system with the intersection coupling constant equal to zero ($k=0)$ has an even higher degeneracy of the vacuum state, $ 2^{3 N}$, because now even the intersecting magnetic sheets  on which the spins are oriented in any of the two directions, represent one of the vacuum configurations \cite{Savvidy:1993sr, Savvidy:1994sc,Pietig:1996xj, Pietig:1997va}.  In recent publications this symmetry was referred to as the subsystem symmetry \cite{Vijay:2016phm}. This high symmetry allows to construct dual representations of these systems  in various  dimensions \cite{Savvidy:1993sr, Savvidy:1994sc,Savvidy:1994tf}. They have rich physical properties, including the glass behaviour  \cite{Sherrington:1975zz, 2010PhRvB..81r4303C, Lipowski:1998yc}  and exotic fracton excitations \cite{Vijay:2016phm}.
 
\section{\it Acknowledgement } 
 
I would like to thank Konstantin Savvidy for stimulating discussion.

\bibliographystyle{elsarticle-num}
\bibliography{magnetic}

\begin{thebibliography}{10}
\expandafter\ifx\csname url\endcsname\relax
  \def\url#1{\texttt{#1}}\fi
\expandafter\ifx\csname urlprefix\endcsname\relax\def\urlprefix{URL }\fi
\expandafter\ifx\csname href\endcsname\relax
  \def\href#1#2{#2} \def\path#1{#1}\fi

\bibitem{Batalin:1976uv}
I.~A. Batalin, S.~G. Matinyan, G.~K. Savvidy, {Vacuum Polarization by a
  Source-Free Gauge Field}, Sov. J. Nucl. Phys. 26 (1977) 214.

\bibitem{Savvidy:1977as}
G.~K. Savvidy, {Infrared Instability of the Vacuum State of Gauge Theories and
  Asymptotic Freedom}, Phys. Lett. B 71 (1977) 133--134.
\newblock \href {http://dx.doi.org/10.1016/0370-2693(77)90759-6}
  {\path{doi:10.1016/0370-2693(77)90759-6}}.

\bibitem{Matinyan:1976mp}
S.~G. Matinyan, G.~K. Savvidy, {Vacuum Polarization Induced by the Intense
  Gauge Field}, Nucl. Phys. B 134 (1978) 539--545.
\newblock \href {http://dx.doi.org/10.1016/0550-3213(78)90463-7}
  {\path{doi:10.1016/0550-3213(78)90463-7}}.

\bibitem{PhDTheses}
G.~{Savvidy}, {Vacuum Polarisation by Intensive Gauge Fields}, PhD, Yerevan
  Physics Institute, 1977.

\bibitem{tHooft:1976snw}
G.~'t~Hooft, {Computation of the Quantum Effects Due to a Four-Dimensional
  Pseudoparticle}, Phys. Rev. D 14 (1976) 3432--3450, [Erratum: Phys.Rev.D 18,
  2199 (1978)].
\newblock \href {http://dx.doi.org/10.1103/PhysRevD.14.3432}
  {\path{doi:10.1103/PhysRevD.14.3432}}.

\bibitem{Nielsen:1978rm}
N.~K. Nielsen, P.~Olesen, {An Unstable Yang-Mills Field Mode}, Nucl. Phys. B
  144 (1978) 376--396.
\newblock \href {http://dx.doi.org/10.1016/0550-3213(78)90377-2}
  {\path{doi:10.1016/0550-3213(78)90377-2}}.

\bibitem{Skalozub:1978fy}
V.~V. Skalozub, {On Restoration of Spontaneously Broken Symmetry in Magnetic
  Field}, Yad. Fiz. 28 (1978) 228--230.

\bibitem{Ambjorn:1978ff}
J.~Ambjorn, N.~K. Nielsen, P.~Olesen, {A Hidden Higgs Lagrangian in {QCD}},
  Nucl. Phys. B 152 (1979) 75--96.
\newblock \href {http://dx.doi.org/10.1016/0550-3213(79)90080-4}
  {\path{doi:10.1016/0550-3213(79)90080-4}}.

\bibitem{Nielsen:1978zg}
H.~B. Nielsen, {Approximate {QCD} Lower Bound for the Bag Constant $B$}, Phys.
  Lett. B 80 (1978) 133--137.
\newblock \href {http://dx.doi.org/10.1016/0370-2693(78)90326-X}
  {\path{doi:10.1016/0370-2693(78)90326-X}}.

\bibitem{Ambjorn:1980ms}
J.~Ambjorn, P.~Olesen, {A Color Magnetic Vortex Condensate in QCD}, Nucl. Phys.
  B 170 (1980) 265--282.
\newblock \href {http://dx.doi.org/10.1016/0550-3213(80)90150-9}
  {\path{doi:10.1016/0550-3213(80)90150-9}}.

\bibitem{Leutwyler:1980ev}
H.~Leutwyler, {Vacuum Fluctuations Surrounding Soft Gluon Fields}, Phys. Lett.
  B 96 (1980) 154--158.
\newblock \href {http://dx.doi.org/10.1016/0370-2693(80)90234-8}
  {\path{doi:10.1016/0370-2693(80)90234-8}}.

\bibitem{Leutwyler:1980ma}
H.~Leutwyler, {Constant Gauge Fields and their Quantum Fluctuations}, Nucl.
  Phys. B 179 (1981) 129--170.
\newblock \href {http://dx.doi.org/10.1016/0550-3213(81)90252-2}
  {\path{doi:10.1016/0550-3213(81)90252-2}}.

\bibitem{Minkowski:1981ma}
P.~Minkowski, {On the Ground State Expectation Value of the Field Strength
  Bilinear in Gauge Theories and Constant Classical Fields}, Nucl. Phys. B 177
  (1981) 203--217.
\newblock \href {http://dx.doi.org/10.1016/0550-3213(81)90388-6}
  {\path{doi:10.1016/0550-3213(81)90388-6}}.

\bibitem{Flory:1983td}
C.~A. Flory, {Covariant Constant Chromomagnetic Fields And Elimination Of The
  One Loop Instabilities}, SLAC-PUB-3244.

\bibitem{Faddeev:2001dda}
L.~D. Faddeev, A.~J. Niemi, {Aspects of Electric magnetic duality in SU(2)
  Yang-Mills theory}, Phys. Lett. B 525 (2002) 195--200.
\newblock \href {http://arxiv.org/abs/hep-th/0101078}
  {\path{arXiv:hep-th/0101078}}, \href
  {http://dx.doi.org/10.1016/S0370-2693(01)01432-0}
  {\path{doi:10.1016/S0370-2693(01)01432-0}}.

\bibitem{Savvidy:2019grj}
G.~Savvidy, {From Heisenberg\textendash{}Euler Lagrangian to the discovery of
  Chromomagnetic Gluon Condensation}, Eur. Phys. J. C 80~(2) (2020) 165.
\newblock \href {http://arxiv.org/abs/1910.00654} {\path{arXiv:1910.00654}},
  \href {http://dx.doi.org/10.1140/epjc/s10052-020-7711-6}
  {\path{doi:10.1140/epjc/s10052-020-7711-6}}.

\bibitem{Pimentel:2018nkl}
G.~L. Pimentel, A.~M. Polyakov, G.~M. Tarnopolsky, {Vacua on the Brink of
  Decay}, Rev. Math. Phys. 30~(07) (2018) 1840013.
\newblock \href {http://arxiv.org/abs/1803.09168} {\path{arXiv:1803.09168}},
  \href {http://dx.doi.org/10.1142/9789813233867_0020}
  {\path{doi:10.1142/9789813233867_0020}}.

\bibitem{Parthasarathy:1983ck}
R.~Parthasarathy, M.~Singer, K.~S. Viswanathan, {The Ground State Of An Su(2)
  Gauge Theory In A Nonabelian Background Field}, Can. J. Phys. 61 (1983)
  1442--1447.
\newblock \href {http://dx.doi.org/10.1139/p83-185}
  {\path{doi:10.1139/p83-185}}.

\bibitem{Kay:1983an}
D.~Kay, R.~Parthasarathy, K.~S. Viswanathan, {Constant Selfdual Abelian Gauge
  Fields And Fermions In Su(2) Gauge Theory}, Phys. Rev. D 28 (1983) 3116.
\newblock \href {http://dx.doi.org/10.1103/PhysRevD.28.3116}
  {\path{doi:10.1103/PhysRevD.28.3116}}.

\bibitem{Kay1983}
D.~{Kay}, {Unstable modes, zero modes, and phase transitions in QCD}, PhD,Simon
  Fraser University, 1985.

\bibitem{Dittrich:1983ej}
W.~Dittrich, M.~Reuter, {Effective {QCD} Lagrangian With Zeta Function
  Regularization}, Phys. Lett. B 128 (1983) 321--326.
\newblock \href {http://dx.doi.org/10.1016/0370-2693(83)90268-X}
  {\path{doi:10.1016/0370-2693(83)90268-X}}.

\bibitem{Zwanziger:1982na}
D.~Zwanziger, {Nonperturbative Modification of the Faddeev-popov Formula and
  Banishment of the Naive Vacuum}, Nucl. Phys. B 209 (1982) 336--348.
\newblock \href {http://dx.doi.org/10.1016/0550-3213(82)90260-7}
  {\path{doi:10.1016/0550-3213(82)90260-7}}.

\bibitem{Kay:2005wm}
D.~Kay, A.~Kumar, R.~Parthasarathy, {Savvidy vacuum in SU(2) Yang-Mills
  theory}, Mod. Phys. Lett. A 20 (2005) 1655--1662.
\newblock \href {http://dx.doi.org/10.1142/S0217732305017913}
  {\path{doi:10.1142/S0217732305017913}}.

\bibitem{Kondo:2013aga}
K.-I. Kondo, {Stability of magnetic condensation and mass generation for
  confinement in SU(2) Yang-Mills theory}, PoS QCD-TNT-III (2013) 020.
\newblock \href {http://arxiv.org/abs/1312.0053} {\path{arXiv:1312.0053}},
  \href {http://dx.doi.org/10.22323/1.193.0020}
  {\path{doi:10.22323/1.193.0020}}.

\bibitem{Cho:2004qf}
Y.~M. Cho, M.~L. Walker, {Stability of monopole condensation in SU(2) QCD},
  Mod. Phys. Lett. A 19 (2004) 2707--2716.
\newblock \href {http://dx.doi.org/10.1142/S0217732304015750}
  {\path{doi:10.1142/S0217732304015750}}.

\bibitem{Pak:2020izt}
D.~G. Pak, R.-G. Cai, T.~Tsukioka, P.~Zhang, Y.-F. Zhou, {Color confinement and
  color singlet structure of quantum states in Yang-Mills theory}\href
  {http://arxiv.org/abs/2011.02926} {\path{arXiv:2011.02926}}.

\bibitem{Pak:2020obo}
D.~G. Pak, T.~Tsukioka, {Color structure of quantum $SU(N)$ Yang-Mills
  theory}\href {http://arxiv.org/abs/2012.11496} {\path{arXiv:2012.11496}}.

\bibitem{Pak:2017skw}
D.~G. Pak, B.-H. Lee, Y.~Kim, T.~Tsukioka, P.~M. Zhang, {On microscopic
  structure of the QCD vacuum}, Phys. Lett. B 780 (2018) 479--484.
\newblock \href {http://arxiv.org/abs/1703.09635} {\path{arXiv:1703.09635}},
  \href {http://dx.doi.org/10.1016/j.physletb.2018.03.040}
  {\path{doi:10.1016/j.physletb.2018.03.040}}.

\bibitem{Pak:2020fkt}
D.~G. Pak, R.-G. Cai, T.~Tsukioka, P.~Zhang, Y.-F. Zhou, {Inherent color
  symmetry in quantum Yang-Mills theory}, Phys. Lett. B 839 (2023) 137804.
\newblock \href {http://arxiv.org/abs/2009.13938} {\path{arXiv:2009.13938}},
  \href {http://dx.doi.org/10.1016/j.physletb.2023.137804}
  {\path{doi:10.1016/j.physletb.2023.137804}}.

\bibitem{Baseian:1979zx}
G.~Z. Baseian, S.~G. Matinyan, G.~K. Savvidy, {Nonlinear Plane Waves In
  Massless Yang-Mills Theory. (In Russian)}, Pisma Zh. Eksp. Teor. Fiz. 29
  (1979) 641--644.

\bibitem{Savvidy:2022jcr}
G.~Savvidy, {Stability of Yang Mills vacuum state}, Nucl. Phys. B 990 (2023)
  116187.
\newblock \href {http://arxiv.org/abs/2203.14656} {\path{arXiv:2203.14656}},
  \href {http://dx.doi.org/10.1016/j.nuclphysb.2023.116187}
  {\path{doi:10.1016/j.nuclphysb.2023.116187}}.

\bibitem{Savvidy:2023kmx}
G.~Savvidy, {Yang\textendash{}Mills effective Lagrangian \textemdash{}
  Contribution of Leutwyler zero mode chromons}, Mod. Phys. Lett. A 38~(06)
  (2023) 2350042.
\newblock \href {http://arxiv.org/abs/2304.01164} {\path{arXiv:2304.01164}},
  \href {http://dx.doi.org/10.1142/S0217732323500426}
  {\path{doi:10.1142/S0217732323500426}}.

\bibitem{Savvidy:2023kft}
G.~Savvidy, {On the stability of Yang-Mills vacuum}, Phys. Lett. B 844 (2023)
  138082.
\newblock \href {http://dx.doi.org/10.1016/j.physletb.2023.138082}
  {\path{doi:10.1016/j.physletb.2023.138082}}.

\bibitem{Savvidy:1982wx}
G.~K. Savvidy, {Yang-Mills Classical Mechanics As A Kolmogorov K System}, Phys.
  Lett. B 130 (1983) 303--307.
\newblock \href {http://dx.doi.org/10.1016/0370-2693(83)91146-2}
  {\path{doi:10.1016/0370-2693(83)91146-2}}.

\bibitem{Savvidy:1982jk}
G.~K. Savvidy, {Classical and Quantum Mechanics of Nonabelian Gauge Fields},
  Nucl. Phys. B 246 (1984) 302--334.
\newblock \href {http://dx.doi.org/10.1016/0550-3213(84)90298-0}
  {\path{doi:10.1016/0550-3213(84)90298-0}}.

\bibitem{Matinyan:1981ys}
S.~G. Matinyan, G.~K. Savvidy, N.~G. Ter-Arutunian~Savvidy, {Stochasticity of
  Classical {Yang-Mills} Mechanics and Its Elimination by Higgs Mechanism. (In
  Russian)}, JETP Lett. 34 (1981) 590--593.

\bibitem{Matinyan:1981dj}
S.~G. Matinyan, G.~K. Savvidy, N.~G. Ter-Arutunian~Savvidy, {Classical
  Yang-Mills Mechanics. Nonlinear Color Oscillations}, Sov. Phys. JETP 53
  (1981) 421--425.

\bibitem{Banks:1996vh}
T.~Banks, W.~Fischler, S.~H. Shenker, L.~Susskind, {M theory as a matrix model:
  A Conjecture}, Phys. Rev. D 55 (1997) 5112--5128.
\newblock \href {http://arxiv.org/abs/hep-th/9610043}
  {\path{arXiv:hep-th/9610043}}, \href
  {http://dx.doi.org/10.1103/PhysRevD.55.5112}
  {\path{doi:10.1103/PhysRevD.55.5112}}.

\bibitem{Savvidy:2020mco}
G.~Savvidy, {Maximally chaotic dynamical systems}, Annals Phys. 421 (2020)
  168274.
\newblock \href {http://dx.doi.org/10.1016/j.aop.2020.168274}
  {\path{doi:10.1016/j.aop.2020.168274}}.

\bibitem{Nielsen:1979vb}
H.~B. Nielsen, P.~Olesen, {Quark Confinement In A Random Color Magnetic Ether}.

\bibitem{Anous:2017mwr}
T.~Anous, C.~Cogburn, {Mini-BFSS matrix model in silico}, Phys. Rev. D 100~(6)
  (2019) 066023.
\newblock \href {http://arxiv.org/abs/1701.07511} {\path{arXiv:1701.07511}},
  \href {http://dx.doi.org/10.1103/PhysRevD.100.066023}
  {\path{doi:10.1103/PhysRevD.100.066023}}.

\bibitem{Cho:1979nv}
Y.~M. Cho, {A Restricted Gauge Theory}, Phys. Rev. D 21 (1980) 1080.
\newblock \href {http://dx.doi.org/10.1103/PhysRevD.21.1080}
  {\path{doi:10.1103/PhysRevD.21.1080}}.

\bibitem{Kim:2016xdn}
Y.~Kim, B.-H. Lee, D.~G. Pak, C.~Park, T.~Tsukioka, {Quantum stability of
  nonlinear wave type solutions with intrinsic mass parameter in QCD}, Phys.
  Rev. D 96~(5) (2017) 054025.
\newblock \href {http://arxiv.org/abs/1607.02083} {\path{arXiv:1607.02083}},
  \href {http://dx.doi.org/10.1103/PhysRevD.96.054025}
  {\path{doi:10.1103/PhysRevD.96.054025}}.

\bibitem{Milshtein:1983th}
A.~I. Milshtein, Y.~F. Pinelis, {Properties of the Photon Polarization Operator
  in a Long Wave Vacuum Field in {QCD}}, Phys. Lett. B 137 (1984) 235.
\newblock \href {http://dx.doi.org/10.1016/0370-2693(84)90236-3}
  {\path{doi:10.1016/0370-2693(84)90236-3}}.

\bibitem{Olesen:1981zp}
P.~Olesen, {Confinement and Random Fields}, Nucl. Phys. B 200 (1982) 381--390.
\newblock \href {http://dx.doi.org/10.1016/0550-3213(82)90094-3}
  {\path{doi:10.1016/0550-3213(82)90094-3}}.

\bibitem{Apenko:1982tj}
S.~Apenko, D.~Kirzhnits, Y.~Lozovik, {Dynamic Chaos, Anderson Localization, And
  Confinement}, JETP Lett. 36 (1982) 213--215.

\bibitem{Reuter:1994yq}
M.~Reuter, C.~Wetterich, {Search for the QCD ground state}, Phys. Lett. B 334
  (1994) 412--419.
\newblock \href {http://arxiv.org/abs/hep-ph/9405300}
  {\path{arXiv:hep-ph/9405300}}, \href
  {http://dx.doi.org/10.1016/0370-2693(94)90707-2}
  {\path{doi:10.1016/0370-2693(94)90707-2}}.

\bibitem{Reuter:1997gx}
M.~Reuter, C.~Wetterich, {Gluon condensation in nonperturbative flow
  equations}, Phys. Rev. D 56 (1997) 7893--7916.
\newblock \href {http://arxiv.org/abs/hep-th/9708051}
  {\path{arXiv:hep-th/9708051}}, \href
  {http://dx.doi.org/10.1103/PhysRevD.56.7893}
  {\path{doi:10.1103/PhysRevD.56.7893}}.

\bibitem{Reuter:1994zn}
M.~Reuter, C.~Wetterich, {Indications for gluon condensation for
  nonperturbative flow equations}\href {http://arxiv.org/abs/hep-th/9411227}
  {\path{arXiv:hep-th/9411227}}.

\bibitem{Wu:1975vq}
T.~T. Wu, C.-N. Yang, {Some Remarks About Unquantized Nonabelian Gauge Fields},
  Phys. Rev. D 12 (1975) 3843--3844.
\newblock \href {http://dx.doi.org/10.1103/PhysRevD.12.3843}
  {\path{doi:10.1103/PhysRevD.12.3843}}.

\bibitem{Wu:1967vp}
T.~T. Wu, C.-N. Yang, {Some Solutions Of The Classical Isotopic Gauge Field
  Equations}, PRINT-67-2362.

\bibitem{Nielsen:1978tr}
H.~B. Nielsen, M.~Ninomiya, {A Bound on Bag Constant and Nielsen-Olesen
  Unstable Mode in QCD}, Nucl. Phys. B 156 (1979) 1--28.
\newblock \href {http://dx.doi.org/10.1016/0550-3213(79)90490-5}
  {\path{doi:10.1016/0550-3213(79)90490-5}}.

\bibitem{Nielsen:1979xu}
H.~B. Nielsen, P.~Olesen, {A Quantum Liquid Model for the QCD Vacuum: Gauge and
  Rotational Invariance of Domained and Quantized Homogeneous Color Fields},
  Nucl. Phys. B 160 (1979) 380--396.
\newblock \href {http://dx.doi.org/10.1016/0550-3213(79)90065-8}
  {\path{doi:10.1016/0550-3213(79)90065-8}}.

\bibitem{Ambjorn:1979xi}
J.~Ambjorn, P.~Olesen, {On the Formation of a Random Color Magnetic Quantum
  Liquid in QCD}, Nucl. Phys. B 170 (1980) 60--78.
\newblock \href {http://dx.doi.org/10.1016/0550-3213(80)90476-9}
  {\path{doi:10.1016/0550-3213(80)90476-9}}.

\bibitem{Brown:1975bc}
M.~R. Brown, M.~J. Duff, {Exact Results for Effective Lagrangians}, Phys. Rev.
  D 11 (1975) 2124--2135.
\newblock \href {http://dx.doi.org/10.1103/PhysRevD.11.2124}
  {\path{doi:10.1103/PhysRevD.11.2124}}.

\bibitem{Duff:1975ue}
M.~J. Duff, M.~Ramon-Medrano, {On the Effective Lagrangian for the Yang-Mills
  Field}, Phys. Rev. D 12 (1975) 3357.
\newblock \href {http://dx.doi.org/10.1103/PhysRevD.12.3357}
  {\path{doi:10.1103/PhysRevD.12.3357}}.

\bibitem{Batalin:1979jh}
I.~A. Batalin, G.~K. Savvidy, {On Gauge Invariance Of Effective Action On
  Precise Sourceless Extremal}, Izv. Akad. Nauk Arm. SSR Fiz. 15 (1980) 3--8.

\bibitem{Cho:1980nx}
Y.~M. Cho, {Extended Gauge Theory and Its Mass Spectrum}, Phys. Rev. D 23
  (1981) 2415.
\newblock \href {http://dx.doi.org/10.1103/PhysRevD.23.2415}
  {\path{doi:10.1103/PhysRevD.23.2415}}.

\bibitem{Cho:2010zzb}
Y.~M. Cho, {QCD effective action and stability of magnetic condensation}, Nucl.
  Phys. A 844 (2010) 120C--137C.
\newblock \href {http://dx.doi.org/10.1016/j.nuclphysa.2010.05.023}
  {\path{doi:10.1016/j.nuclphysa.2010.05.023}}.

\bibitem{tHooft:1974kcl}
G.~'t~Hooft, {Magnetic Monopoles in Unified Gauge Theories}, Nucl. Phys. B 79
  (1974) 276--284.
\newblock \href {http://dx.doi.org/10.1016/0550-3213(74)90486-6}
  {\path{doi:10.1016/0550-3213(74)90486-6}}.

\bibitem{Corrigan:1975zxj}
E.~Corrigan, D.~I. Olive, D.~B. Fairlie, J.~Nuyts, {Magnetic Monopoles in SU(3)
  Gauge Theories}, Nucl. Phys. B 106 (1976) 475--492.
\newblock \href {http://dx.doi.org/10.1016/0550-3213(76)90173-5}
  {\path{doi:10.1016/0550-3213(76)90173-5}}.

\bibitem{Biran:1987ae}
B.~Biran, E.~G.~F. Floratos, G.~K. Savvidy, {The Selfdual Closed Bosonic
  Membranes}, Phys. Lett. B 198 (1987) 329--332.
\newblock \href {http://dx.doi.org/10.1016/0370-2693(87)90673-3}
  {\path{doi:10.1016/0370-2693(87)90673-3}}.

\bibitem{Wu:1975es}
T.~T. Wu, C.~N. Yang, {Concept of Nonintegrable Phase Factors and Global
  Formulation of Gauge Fields}, Phys. Rev. D 12 (1975) 3845--3857.
\newblock \href {http://dx.doi.org/10.1103/PhysRevD.12.3845}
  {\path{doi:10.1103/PhysRevD.12.3845}}.

\bibitem{Goddard:1976qe}
P.~Goddard, J.~Nuyts, D.~I. Olive, {Gauge Theories and Magnetic Charge}, Nucl.
  Phys. B 125 (1977) 1--28.
\newblock \href {http://dx.doi.org/10.1016/0550-3213(77)90221-8}
  {\path{doi:10.1016/0550-3213(77)90221-8}}.

\bibitem{Hopf:1931}
H.~Hopf, {Uber die Abbildungen der dreidimensionalen Sphare auf die
  Kugelflache}, Math. Ann. 104 (1931) 637--665.
\newblock \href {http://dx.doi.org/doi.org/10.1007/BF01457962}
  {\path{doi:doi.org/10.1007/BF01457962}}.

\bibitem{Whitehead:1947}
J.~H.~C. Whitehead, {An Expression of Hopf's Invariant as an Integral},
  Proceedings of the National Academy of Sciences 33 (1947) 117--123.
\newblock \href {http://dx.doi.org/doi:10.1073/pnas.33.5.117}
  {\path{doi:doi:10.1073/pnas.33.5.117}}.

\bibitem{cmp/1104178138}
E.~Witten, {Quantum field theory and the Jones polynomial}, Communications in
  Mathematical Physics 121~(3) (1989) 351 -- 399.

\bibitem{Moffatt:1969}
M.~H.K., {The degree of knottedness of tangled vortex lines}, J.Fluid Mech. 35
  (1969) 117--129.
\newblock \href {http://dx.doi.org/doi.org/10.1017/S0022112069000991}
  {\path{doi:doi.org/10.1017/S0022112069000991}}.

\bibitem{Moffatt:1991}
M.~H.K., {Relaxation under topological constraints}, in: C.~P. Moffatt H.~K.,
  Zaslavsky G.~M., T.~M. (Eds.), Topological Aspects of the Dynamics of Fluids
  and Plasmas, Vol. 218 of NATO ASI E, Kluwer, Proceedings of the Program of
  the Institute for Theoretical Physics, UCSB, 1991.

\bibitem{Antoniadis:2012ep}
I.~Antoniadis, G.~Savvidy, {New gauge anomalies and topological invariants in
  various dimensions}, Eur. Phys. J. C 72 (2012) 2140.
\newblock \href {http://arxiv.org/abs/1205.0027} {\path{arXiv:1205.0027}},
  \href {http://dx.doi.org/10.1140/epjc/s10052-012-2140-9}
  {\path{doi:10.1140/epjc/s10052-012-2140-9}}.

\bibitem{Jackiw:1976pf}
R.~Jackiw, C.~Rebbi, {Vacuum Periodicity in a Yang-Mills Quantum Theory}, Phys.
  Rev. Lett. 37 (1976) 172--175.
\newblock \href {http://dx.doi.org/10.1103/PhysRevLett.37.172}
  {\path{doi:10.1103/PhysRevLett.37.172}}.

\bibitem{Callan:1976je}
C.~G. Callan, Jr., R.~F. Dashen, D.~J. Gross, {The Structure of the Gauge
  Theory Vacuum}, Phys. Lett. B 63 (1976) 334--340.
\newblock \href {http://dx.doi.org/10.1016/0370-2693(76)90277-X}
  {\path{doi:10.1016/0370-2693(76)90277-X}}.

\bibitem{tHooft:1981bkw}
G.~'t~Hooft, {Topology of the Gauge Condition and New Confinement Phases in
  Nonabelian Gauge Theories}, Nucl. Phys. B 190 (1981) 455--478.
\newblock \href {http://dx.doi.org/10.1016/0550-3213(81)90442-9}
  {\path{doi:10.1016/0550-3213(81)90442-9}}.

\bibitem{Kapustin:2005py}
A.~Kapustin, {Wilson-'t Hooft operators in four-dimensional gauge theories and
  S-duality}, Phys. Rev. D 74 (2006) 025005.
\newblock \href {http://arxiv.org/abs/hep-th/0501015}
  {\path{arXiv:hep-th/0501015}}, \href
  {http://dx.doi.org/10.1103/PhysRevD.74.025005}
  {\path{doi:10.1103/PhysRevD.74.025005}}.

\bibitem{Kapustin:2006pk}
A.~Kapustin, E.~Witten, {Electric-Magnetic Duality And The Geometric Langlands
  Program}, Commun. Num. Theor. Phys. 1 (2007) 1--236.
\newblock \href {http://arxiv.org/abs/hep-th/0604151}
  {\path{arXiv:hep-th/0604151}}, \href
  {http://dx.doi.org/10.4310/CNTP.2007.v1.n1.a1}
  {\path{doi:10.4310/CNTP.2007.v1.n1.a1}}.

\bibitem{Gukov:2006jk}
S.~Gukov, E.~Witten, {Gauge Theory, Ramification, And The Geometric Langlands
  Program}\href {http://arxiv.org/abs/hep-th/0612073}
  {\path{arXiv:hep-th/0612073}}.

\bibitem{Gukov:2008sn}
S.~Gukov, E.~Witten, {Rigid Surface Operators}, Adv. Theor. Math. Phys. 14~(1)
  (2010) 87--178.
\newblock \href {http://arxiv.org/abs/0804.1561} {\path{arXiv:0804.1561}},
  \href {http://dx.doi.org/10.4310/ATMP.2010.v14.n1.a3}
  {\path{doi:10.4310/ATMP.2010.v14.n1.a3}}.

\bibitem{Wegner:1971app}
F.~J. Wegner, {Duality in Generalized Ising Models and Phase Transitions
  Without Local Order Parameters}, J. Math. Phys. 12 (1971) 2259--2272.
\newblock \href {http://dx.doi.org/10.1063/1.1665530}
  {\path{doi:10.1063/1.1665530}}.

\bibitem{Savvidy:1993ej}
G.~K. Savvidy, F.~J. Wegner, {Geometrical string and spin systems}, Nucl. Phys.
  B 413 (1994) 605--613.
\newblock \href {http://arxiv.org/abs/hep-th/9308094}
  {\path{arXiv:hep-th/9308094}}, \href
  {http://dx.doi.org/10.1016/0550-3213(94)90003-5}
  {\path{doi:10.1016/0550-3213(94)90003-5}}.

\bibitem{Savvidy:2000zq}
G.~K. Savvidy, {The System with exponentially degenerate vacuum state}\href
  {http://arxiv.org/abs/cond-mat/0003220} {\path{arXiv:cond-mat/0003220}}.

\bibitem{Savvidy:1993sr}
G.~K. Savvidy, K.~G. Savvidy, {Self-avoiding gonihedric string and spin
  systems}, Phys. Lett. B 324 (1994) 72--77.
\newblock \href {http://arxiv.org/abs/hep-lat/9311026}
  {\path{arXiv:hep-lat/9311026}}, \href
  {http://dx.doi.org/10.1016/0370-2693(94)00114-6}
  {\path{doi:10.1016/0370-2693(94)00114-6}}.

\bibitem{Savvidy:1994sc}
G.~K. Savvidy, K.~G. Savvidy, {Interaction hierarchy}, Phys. Lett. B 337 (1994)
  333--339.
\newblock \href {http://arxiv.org/abs/hep-th/9409030}
  {\path{arXiv:hep-th/9409030}}, \href
  {http://dx.doi.org/10.1016/0370-2693(94)90984-9}
  {\path{doi:10.1016/0370-2693(94)90984-9}}.

\bibitem{Pietig:1996xj}
R.~Pietig, F.~J. Wegner, {Phase transition in lattice surface systems with
  gonihedric action}, Nucl. Phys. B 466 (1996) 513--526.
\newblock \href {http://arxiv.org/abs/hep-lat/9604013}
  {\path{arXiv:hep-lat/9604013}}, \href
  {http://dx.doi.org/10.1016/0550-3213(96)00072-7}
  {\path{doi:10.1016/0550-3213(96)00072-7}}.

\bibitem{Pietig:1997va}
R.~Pietig, F.~J. Wegner, {Low temperature expansion of the gonihedric Ising
  model}, Nucl. Phys. B 525 (1998) 549--570.
\newblock \href {http://arxiv.org/abs/hep-lat/9712002}
  {\path{arXiv:hep-lat/9712002}}, \href
  {http://dx.doi.org/10.1016/S0550-3213(98)00342-3}
  {\path{doi:10.1016/S0550-3213(98)00342-3}}.

\bibitem{Vijay:2016phm}
S.~Vijay, J.~Haah, L.~Fu, {Fracton Topological Order, Generalized Lattice Gauge
  Theory and Duality}, Phys. Rev. B 94~(23) (2016) 235157.
\newblock \href {http://arxiv.org/abs/1603.04442} {\path{arXiv:1603.04442}},
  \href {http://dx.doi.org/10.1103/PhysRevB.94.235157}
  {\path{doi:10.1103/PhysRevB.94.235157}}.

\bibitem{Savvidy:1994tf}
G.~K. Savvidy, K.~G. Savvidy, F.~J. Wegner, {Geometrical string and dual spin
  systems}, Nucl. Phys. B 443 (1995) 565--580.
\newblock \href {http://arxiv.org/abs/hep-th/9503213}
  {\path{arXiv:hep-th/9503213}}, \href
  {http://dx.doi.org/10.1016/0550-3213(95)00151-H}
  {\path{doi:10.1016/0550-3213(95)00151-H}}.

\bibitem{Sherrington:1975zz}
D.~Sherrington, S.~Kirkpatrick, {Solvable Model of a Spin-Glass}, Phys. Rev.
  Lett. 35 (1975) 1792--1796.
\newblock \href {http://dx.doi.org/10.1103/PhysRevLett.35.1792}
  {\path{doi:10.1103/PhysRevLett.35.1792}}.

\bibitem{2010PhRvB..81r4303C}
C.~{Castelnovo}, C.~{Chamon}, D.~{Sherrington}, {Quantum mechanical and
  information theoretic view on classical glass transitions}, PRB 81~(18)
  (2010) 184303.
\newblock \href {http://arxiv.org/abs/1003.3832} {\path{arXiv:1003.3832}},
  \href {http://dx.doi.org/10.1103/PhysRevB.81.184303}
  {\path{doi:10.1103/PhysRevB.81.184303}}.

\bibitem{Lipowski:1998yc}
A.~Lipowski, D.~Johnston, {Glassy transition and metastability in four spin
  Ising model}, J. Phys. A 33 (2000) 4451--4460.
\newblock \href {http://arxiv.org/abs/cond-mat/9812098}
  {\path{arXiv:cond-mat/9812098}}, \href
  {http://dx.doi.org/10.1088/0305-4470/33/24/304}
  {\path{doi:10.1088/0305-4470/33/24/304}}.

\end{thebibliography}

\end{document}